\documentclass [12pt]{article}
\def\maj#1{\ifmmode\mbox{\usefont{U}{msb}{m}{n}#1}\else{\usefont{U}{msb}{m}{n}#1}\fi}
\def\v#1{\mathbf{#1}}
\usepackage{graphics,epsfig,graphicx}
\usepackage{color}

\begin{document}

\title{\textbf{Shiva diagrams for composite-boson many-body effects :
How they work}}
\author{M. Combescot and O. Betbeder-Matibet
\\ \small{\textit{Institut des NanoSciences de Paris,}}\\
\small{\textit{Universit\'e Pierre et Marie Curie-Paris 6, 
Universit\'{e} Denis Diderot-Paris 7, CNRS, UMR 7588,}}\\
\small{\textit{Campus Boucicaut, 140 rue de Lourmel, 75015 Paris}}}
\date{}
\maketitle

\begin{abstract}
The purpose of this paper is to show how the diagrammatic expansion
in fermion exchanges of scalar products of $N$-composite-boson
(``coboson'') states can be obtained in a practical way. The hard
algebra on which this expansion is based, will be given in an independent publication.

Due to the composite nature of the particles, the scalar products
of $N$-coboson states do not reduce to a set of Kronecker symbols, as
for elementary bosons, but contain subtle exchange terms between two or
more cobosons. These terms originate from Pauli exclusion between the
fermionic components of the particles. While our many-body
theory for composite bosons leads to write these scalar products as
complicated sums of products of ``Pauli scatterings'' between
\emph{two} cobosons, they in fact correspond to fermion exchanges
between any number P of quantum particles, with
$2 \leq P\leq N$. These $P$-body exchanges are nicely represented by the
so-called ``Shiva diagrams'', which are topologically different from
Feynman diagrams, due to the intrinsic many-body nature of Pauli
exclusion from which they originate. These Shiva diagrams in fact
constitute the novel part of our composite-exciton many-body theory
which was up to now missing to get its full
diagrammatic representation. Using them, we can now ``see'' through
diagrams the physics of any quantity in which enters $N$ interacting
excitons --- or more generally $N$ composite bosons ---, with fermion
exchanges included in an
\emph{exact} --- and transparent --- way.
\end{abstract}

PACS.: 

71.35.-y Excitons and related phenomena 

05.30.Ch Quantum ensemble theory 

05.30.Jp Boson systems

\newpage

\section{Introduction}

Over the last few years, we have developed a new many-body procedure
[1-3] able to treat interactions between composite excitons, without, at
any stage, replacing them by elementary bosons. The challenge was to
find an exact but tractable way to take care of Pauli exclusion between
the fermionic components of the excitons. This Pauli exclusion is
usually included through carrier exchanges in the
Coulomb scattering of two excitons. Beside the fact that the
effective Coulomb scattering between two bosonized excitons used up to now should
have been rejected long ago because it generates an unacceptable
non-hermiticity in the effective Hamiltonian of boson-excitons [4,5], --- this
non-hermiticity being not the signature of any novel relaxation appearing in the
problem, but the bare consequence of an inconsistent procedure, --- it is fully
unsatisfactory to include Pauli exclusion, which is
$N$-body by essence, through effective Coulomb scatterings between
\emph{two} excitons only : Terms in which the carriers of more than two
excitons are mixed by exchange also exist, as well as terms in which
the excitons see each other, \emph{i.\ e.}, ``interact'', just by
carrier exchange, \emph{without} any Coulomb process. Among its
accomplishments, our new many-body theory for composite excitons allows
us to produce these pure exchange terms in a natural way. It is of importance to note
that the ``scatterings'' associated to fermion exchanges being dimensionless by
construction, they cannot appear in an effective Hamiltonian with the interaction
written as a potential between bosons, whatever the bosonization procedure which
produces this potential is, due to a bare dimensional argument. As a main consequence,
our many-body theory rules out all attempts to correctly describe many-body
effects between composite excitons through an effective Hamiltonian, even at the lowest
order in density, since the pure exchange processes are going to be systematically
missed. Let us stress that these pure exchange terms are crucial for all
semiconductor optical nonlinear effects, because they are responsible for
processes which are dominant at large detuning (see the final results of references
[6-9]).

Quite recently [10], we have formally extended this many-body theory
for composite excitons to any type of composite bosons --- ``cobosons'' in short ---
made of two different fermions $\alpha$ and $\beta$, having in mind its possible
extension to the many-body physics of ultracold atomic gases. This
extension is conveniently done by introducing an (arbitrary) orthogonal
basis for free fermion pairs, with a closure relation given by
\begin{equation}
I=\sum_{\v k_\alpha,\v k_\beta}|\v k_\alpha,\v
k_\beta\rangle\,\langle\v k_\beta,\v k_\alpha|\ .
\end{equation}
Any composite-boson state $|i\rangle$ made of \emph{one} fermion $\alpha$
and \emph{one} fermion $\beta$ expands on this basis as
\begin{equation}
|i\rangle=\sum_{\v k_\alpha,\v k_\beta}|\v k_\alpha,\v
k_\beta\rangle\,\langle\v k_\beta,\v k_\alpha|i\rangle\ .
\end{equation}
So that the creation operator of this one-coboson state
$|i\rangle=B_i^\dag|v\rangle$ reads in terms of the creation
operators for free fermions, $|\v k_\alpha,\v k_\beta\rangle=a_{\v
k_\alpha}^\dag b_{\v k_\beta}^\dag|v\rangle$ as
\begin{equation}
B_i^\dag=\sum_{\v k_\alpha,\v k_\beta}\langle\v k_\beta,\v
k_\alpha|i\rangle \,a_{\v k_\alpha}^\dag b_{\v k_\beta}^\dag\ .
\end{equation}
If the composite bosons $|i\rangle$ also form a complete orthogonal set [11] for
one-fermion-pair states, as for $|i\rangle$ being the one-pair eigenstates of the
Hamiltonian, we also have
\begin{equation}
I=\sum_i|i\rangle\,\langle i|\ ,
\end{equation}
so that the free-fermion-pair creation operators can, in the same way, be written in
terms of coboson operators as
\begin{equation}
a_{\v k_\alpha}^\dag b_{\v k_\beta}^\dag=\sum_i\langle i|\v k_\alpha,\v
k_\beta\rangle\,B_i^\dag\ .
\end{equation}
This last equation allows to rewrite any
physical quantity dealing with
$N$ pairs of fermions $(\alpha,\beta)$, in terms of matrix elements
between $N$-coboson states like
\begin{equation}
\langle v|B_{m_N}\cdots B_{m_1}\,f(H)\,B_{i_1}^\dag\cdots B_{i_N}^\dag
|v\rangle\ ,
\end{equation}
(with additional functions of the system Hamiltonian $H$ possibly in other
places than the middle). In
order to calculate these matrix elements, we first push $f(H)$ to the
right, using the commutator $[f(H),B_i^\dag]$ which can be deduced
from
\begin{equation}
[H,B_i^\dag] =E_i\,B_i^\dag+V_i^\dag\ ,
\end{equation}
the above equation being valid when the one-pair state $|i\rangle$ is a one-pair
eigenstate [12] of the Hamiltonian,
$(H-E_i)|i\rangle=0$. The ``creation potential'' $V_i^\dag$ is then
eliminated from the matrix element through
\begin{equation}
[V_i^\dag,B_j^\dag] =
\sum_{mn}\xi\left(^{n\ \,j}_{m\ i}\right)\,B_m^\dag
B_n^\dag\ ,
\end{equation}
where $\xi\left(^{n\ \,j}_{m\ i}\right)$ is the \emph{direct}
interaction scattering between cobosons in states $(i,j)$ ending
in states $(m,n)$. It comes from the interactions which exist between the fermions
$\alpha$ and
$\beta$ of the cobosons $(i,j)$ : This scattering is ``direct'' in the
sense that the cobosons $m$ and $i$ are made with the same fermion
pair, and similarly for $n$ and $j$.

The $f(H)$'s of physical interest are $1/(a-H)$ which appears in
problems dealing with correlation effects or response functions and
$e^{-iHt}$ in problems dealing with time evolution. The precise values
of $[f(H),B_i^\dag]$ in terms of $V_j^\dag$, for these $f(H)$, can be
found in eqs.\ (10-11) of reference [3].

By pushing $f(H)$ to the right, we generate a lot of scatterings
$\xi$. Being direct scatterings between two cobosons by construction,
the corresponding terms can be visualized through diagrams with
scatterings between two coboson lines, as in Fig.1 : The diagrams representing these
direct processes between cobosons are very similar to Feynman diagrams for elementary
quantum particles, like electrons interacting through Coulomb interaction.

The original -- and difficult -- part of problems dealing with
interacting composite bosons is the calculation of the
remaining terms, \emph{i}.\ \emph{e}., terms like the one of eq.\
(6) with
$f(H)$ replaced by 1. The
scalar products of these $N$-coboson states can \emph{a priori} be
calculated by pushing the
$B$'s to the right, through the commutator
\begin{equation}
[B_m,B_i^\dag]=\delta_{m,i}-D_{mi}\ ,
\end{equation}
and by eliminating the deviation-from-boson operator $D_{mi}$ through
[1,10]
\begin{equation}
[D_{mi},B_j^\dag]=\sum_{n}\left[\lambda\left(^{n\ \,j}_{m\
i}\right)+\lambda\left(^{m\ j}_{n\ \,i}\right)\right]\,B_n^\dag\ ,
\end{equation}
which follows from eqs.\ (3,5) and the fact that $\langle
m|i\rangle=\delta_{m,i}$ for cobosons forming an orthonormal set.
The 2-body Pauli scatterings $\lambda\left(^{n\ \,j}_{m\ i}\right)$ which appear in
these manipulations, read in terms of the coboson wave functions $\phi_i(\v r_\alpha,\v
r_\beta)=
\langle\v r_\beta,\v r_\alpha|i\rangle$ as
\begin{equation}
\lambda\left(^{n\ \,j}_{m\ i}\right)=\int d\v r_{\alpha 1}\,d\v
r_{\beta 1}\, d\v r_{\alpha 2}\,d\v r_{\beta 2}\,\phi_m^\ast(\v
r_{\alpha 1},\v r_{\beta 2})\,\phi_n^\ast(\v r_{\alpha 2},\v
r_{\beta 1})\,\phi_i(\v r_{\alpha 1},\v r_{\beta 1})\,\phi_j(\v
r_{\alpha 2},\v r_{\beta 2})\ ,
\end{equation}
the cobosons of the bottom line, $m$ and $i$, by construction having the same
fermion $\alpha$, here located at $\v r_{\alpha 1}$.

The commutators (9,10) make the
$N$-coboson scalar products reading as a sum of products of Pauli
scatterings between \emph{two} cobosons. For $N=2$, these
commutators readily give [1,10] 
\begin{equation}
\langle v|B_mB_nB_i^\dag B_j^\dag|v\rangle=\delta_{m,i}\delta_{n,j}+
\delta_{m,j}\delta_{n,i}-\lambda\left(^{n\ \,j}_{m\
i}\right)-\lambda\left(^{m\ j}_{n\ \,i}\right)\ ,
\end{equation}
which is represented by the diagrams of Fig.2.
For cobosons considered as elementary bosons, the Pauli scatterings
$\lambda$ reduce to zero, so that such a scalar product only
contains the $\delta$ terms, \emph{i.\ e.}, the two first diagrams of
Fig.2.

$N$-coboson scalar products can still be calculated rather
easily for $N=3$ by using the commutators (9,10). We find 
\begin{eqnarray}
\langle v|B_mB_nB_pB_i^\dag B_j^\dag
B_k^\dag|v\rangle=\left[\delta_{m,i}\,\delta_{n,j}\,\delta_{p,k}+\
\mathrm{perm.}\right]-\left[\delta_{m,i}\,\lambda\left(^{p\ k}_{n\ j}
\right)+\ \mathrm{perm.}\right]\nonumber\\
+\sum_{s}\left[\lambda\left(^{p\ k}_{n\ s}\right)\lambda\left(^{s\
\,j}_{m\ i}\right)+\ \mathrm{perm.}\right]\ .
\end{eqnarray}
However, such a
pedestrian procedure becomes totally hopeless when the number of
cobosons $N$ gets large. This is why it is highly necessary to
find a better procedure to calculate these scalar products, if we
want to handle many-body effects between cobosons really, \emph{i.\
e.}, not the ones involving just two or three composite bosons.

\emph{The purpose of this paper is to propose a direct procedure to
construct the diagrammatic expansion in fermion exchanges of the
scalar products of $N$-coboson states and to calculate the
corresponding diagrams in a ``visual'' way, any physical effect
involving $N$ pairs of fermions ultimately reading in terms of these scalar
products.}

Before going further, let us make clear the fact that ``excitons'' are nothing but
particular ``cobosons'', the ones made with electron-hole pairs interacting by Coulomb
potentials. Many-body effects in semiconductors are due to interactions between the $N$
electron-hole pairs contained in the sample, these $N$ pairs being usually called
``$N$ excitons'', unproperly. Indeed, the creation operator of an exciton is a
well defined mathematical object (see eq.\ (3)), so that the
$N$-exciton state $B_{i_1}^\dag\cdots B_{i_N}^\dag|v\rangle$ is \emph{mathematically}
defined without ambiguity. On the opposite, the
excitons are not well defined \emph{physical} objects for systems having more than one
electron-hole pair, since there is no way to identify these objects properly due to the
undistinguishability of the carriers from which they are made. In order to make specific
this difference in vocabulary --- which of course covers a difference in the
understanding --- let us consider the ground state of $N$ electron-hole pairs. This
$N$-pair state is close to the $N$-exciton state
$B_0^{\dag N}|v\rangle$, with $B_0^\dag|v\rangle$ being the one-pair ground state, which
also is the exciton ground state. Due to Pauli and Coulomb scatterings between excitons,
the
$N$-pair ground state also has contributions on other $N$-exciton states like $B_{i\neq
0}^\dag B_{j\neq 0}^\dag B_0^{\dag N-2}|v\rangle$. Consequently, as already
said in connection with eq.\ (6), the physics involving $N$ electron-hole pairs --- 
$N$ fermion pairs in general --- ultimately reads in terms of the $N$-coboson states
considered here.
Their scalar products thus constitute the keys 
to control all many-body effects involving $N$ fermion pairs in the low density
limit.

In usual problems dealing with $N$ cobosons, most of them are
in the same state 0 (often the ground state). A few years ago [13,14],
we have calculated the simplest of these scalar products, namely
\begin{equation}
\langle v|B_0^N\,B_0^{\dag N}|v\rangle=N!\,F_N\ .
\end{equation}
Its calculation turned out to be rather tricky already.
While $F_N$ reduces to 1 if the cobosons are taken as
elementary bosons, we have been surprised to find that, for
composite bosons,
$F_N$ is not a corrective factor of the order of 1, but an
underextensive quantity which decreases exponentially with $N$. In
the case of 3D excitons, $F_N$ precisely reads [13,14] 
\begin{equation}
F_N\simeq\exp N\left[-\frac{33\pi}{4}\eta+\frac{233\pi^2}{6}
\eta^2+\cdots\right]\ ,
\end{equation}
where $\eta=Na_X^3/L^3$ is the exciton density in Bohr radius unit,
so that, although $\eta$ is always small when excitons exist, $N\eta$
can be much larger than 1 if the sample is very large. Actually, in
physical quantities, this underextensive factor
only enters through ratios like
$F_{N-p}/F_N$, so that, as possibly seen from to eq.\ (15), the
correction these ratios induce [14] reduces to $1+O(\eta)$. 

In a work [15] dealing with the Hamiltonian expectation value in the
$N$-ground-state-exciton state --- which is a way to reach the part of the ground state
energy of $N$ electron-hole pairs coming from their interactions treated in the Born
approximation, --- we have been led to calculate two other scalar products
of
$N$-exciton states, with one or two excitons in a state different from 0 on the same
side, namely
$\langle v|B_0^{N-1}B_mB_0^{\dag N}|v\rangle$ and $\langle
v|B_0^{N-2}B_m B_n B_0 ^{\dag N}|v\rangle$. In order to get
them, we first derived recursion relations between these scalar
products for $N$, 
$(N-1)$\ldots excitons in terms of the Pauli scatterings
$\lambda$ between two excitons, using eqs.\ (9,10). These
recursion relations allow us to generate the expansion of these scalar
products in ``Pauli diagrams'', \emph{i}.\ \emph{e}., diagrams
written with $2\times 2$ Pauli scatterings. These diagrams make
appearing a lot of irrelevant intermediate exciton states over which
sums are taken, the final result only depending on the exciton states
involved in the matrix element at hand.

We started to realize that the ``Pauli diagrams'' involving
scatterings between \emph{two} excitons only, were definitely not the
good diagrammatic representation of the $N$-exciton state scalar
products, when we tried to calculate $\langle v|B_0^{N-1}B_mB_i^\dag
B_0^{\dag N-1}|v\rangle$, with an exciton different from 0 on each
side [16]. Indeed, depending on the way we perform the commutations of
the
$B$'s with the $B^\dag$'s, we generate different recursion relations
which give rise to topologically different Pauli diagrams, although
all these diagrams, of course, represent the same quantity. In order
to show the equivalence of all these Pauli diagrams, we were forced to
formally perform the summation over the (irrelevant) intermediate
exciton states generated by the $2\times 2$ Pauli scatterings. This
revealed to us that, although there is not a one-to-one
correspondance, the sets of Pauli diagrams we had
obtained, and which are topologically so different, in fact
correspond to a unique set of new diagrams that we call ``Shiva
diagrams'' [17], in which the intermediate irrelevant exciton states do
not appear anymore. These Shiva diagrams make transparent
the carrier exchanges which take place between two or more of the
excitons appearing in the scalar product of interest. This is after all
rather satisfactory because carrier exchanges can \emph{a priori} exist
between more than just two excitons. It is also clear that multiple
carrier exchanges can be decomposed into a set of $2\times 2$
exchanges, this decomposition being in most cases not unique; this
is why the various representations of a given matrix element in
Pauli diagrams with $2\times 2$ scatterings end by appearing very differently.

These Shiva diagrams turn out to be quite convenient for a
systematic expansion in fermion exchanges of $N$-coboson state
scalar products. By adding to them the direct scatterings
$\xi$ between two coboson lines, which come from the
$H$ contributions to the matrix elements of interest, we end by having
found the full diagrammatic representation of our new many-body theory
for composite bosons, with all possible fermion exchanges included
in an exact --- and transparent --- way.

It is of importance to note that, just from a bare counting of the number of events,
we are led to associate the density expansion of any physical effect involving the
interactions of
$N$ identical composite bosons to Shiva diagrams with an increasing
number of coboson lines : The first order term in density is made of diagrams having two
coboson lines, the second order term is made of diagrams with three lines, and so on,
these  coboson lines being connected by direct interactions between the fermions making
the cobosons and/or by fermion exchanges between these cobosons. This is why, in order
to possibly generate the density expansion of many-body effects between composite bosons,
it is necessary to master the expansion in fermion exchanges of all the possible scalar
products of
$N$-coboson states. The Shiva diagrams we here present appear as highly necessary to
visualize them, due to the extreme complexity of these exchanges when a large number of
cobosons are involved.

In this paper, we 
draw and show how to calculate the Shiva diagrams for six scalar
products of increasing complexity. This appears to us the best way
to make the reader grasping how these new diagrams really work for the
calculations of coboson-state scalar products.

In the various physical effects we have up to now studied, we have had to use
some of these scalar products. The ones of sections 3 and 4 enter the density expansion
of the Hamiltonian expectation value in the state made with $N$
ground state excitons [15]. They also enter the detuning dominant term of the Faraday
rotation produced in photoexcited semiconductors [9]. We have used the scalar products
of section 5 to get the next order term in detuning of this Faraday rotation ---
which comes from processes in which one Coulomb interaction enters [9]. Finally, the
knowledge of the scalar product calculated in section 6 is necessary to obtain the time
evolution of $N$ ground state excitons, in order to get the transition rates and
lifetime of this state as done in ref.\ [18]. In these previous works, the calculation
of the scalar products was somehow hidden in appendices or just not given at all in the
case of letter publications. Since their knowledges are necessary to calculate any other
physical effect involving $N$ fermion pairs we are going to study in the future, it
appears to us as necessary to collect all these results in a single paper, its coherent
presentation making clear the structure of the various exchanges entering these scalar
products.

The precise rules to calculate these Shiva diagrams are based on the recursion
relations which exist between scalar products involving $N$, $N-1$,
$N-2$,Ê\ldots cobosons. The precise algebraic derivation of
these rules turns out to be extremely heavy. This is why we found more
appropriate to concentrate their derivations in a second (highly
technical) paper and to, in this first paper, just leave the description
of
\emph{what has to be done in practice}, if we want to calculate a given
$N$-coboson state scalar product. In reference [16], it is already
possible to find the justification of these
expansions in the case of the scalar products considered in
sections 3 and 6.

\section{Description of Shiva diagrams and rules to get their explicit values}

Before explaining how the diagrammatic expansion in fermion exchanges
of the scalar product of $N$-coboson states can be carried out in a
practical way, let us first introduce the $N$-body
Shiva diagrams which enter these expansions, in an intuitive way.

(i) \emph{In the case of $N=2$ cobosons}, one of the possible fermion
exchanges between cobosons $(i,j)$ shown in Fig.3(a), corresponds
to have $(m,i)$ made with the same fermion $\alpha$ and $(m,j)$
made with the same fermion $\beta$. Actually, this fermion exchange
is topologically equivalent to the diagram shown in Fig.3(b); so that
we can note it in either way,
\begin{equation}
\lambda_2\left(\begin{array}{ll} n&j\\m&i\end{array}\right)=
\lambda_2\left(\begin{array}{ll} m&i\\n&j\end{array}\right)\ .
\end{equation}
By convention, in all the $\lambda_N$'s that we are going to introduce,
the cobosons of the lower lines, here
$(m,i)$ or
$(n,j)$, are made with \emph{the same fermion $\alpha$}. These two
$\lambda_2$'s correspond to the integral given in eq.\ (11), \emph{i.e.}, the
Pauli scattering appearing in the commutator between 2 cobosons given in eq.\ (10).
This can be readily seen from Fig.4(a) and the rules to calculalte
Shiva diagrams that we now give.

(ii) \emph{Rules to calculate Shiva diagrams}

\begin{itemize}

\item Take the wave functions of the ``in'' cobosons, \emph{i.\ e.},
$(i,j)$ in the case of $\lambda_2$, and the complex conjugate of the wave
functions of the ``out'' cobosons, \emph{i.\ e.}, $(m,n)$ in the case of
$\lambda_2$.

\item Write the wave functions of the ``in'' cobosons 1,2,\ldots,$N$ with the
variables
$(\v r_{\alpha_1},\v r_{\beta_1})$, $(\v r_{\alpha_2},\v r_{\beta_2})$, \ldots ,$(\v
r_{\alpha_N},\v r_{\beta_N})$ (see Fig.4).

\item Write the wave functions of the ``out'' cobosons with the variables you read on
the Shiva diagram : In the case of $\lambda_2$, the wave
function of the coboson $m$ appears as $\phi_m^\ast(\v r_{\alpha_1},
\v r_{\beta_2})$, since the coboson $m$ has the same
fermion
$\alpha$ as $i$ and the same fermion $\beta$ as $j$ (see Fig.4).

\item Sum over all variables $(\v r_{\alpha_n},\v r_{\beta_n})$.

\end{itemize}

(iii) \emph{In the case of $N=3$ cobosons}, 
one of the possible fermion exchanges, shown in Fig.3c,
corresponds to have the cobosons $(m,i)$ made with the same fermion
$\alpha$ --- and similarly for $(p,j)$ and $(n,k)$ --- while the
cobosons $(m,j)$ have
the same fermion $\beta$ --- and similarly for $(n,i)$ and $(p,k)$.
This fermion exchange between \emph{three}
cobosons can be represented by one of the \emph{three} topologically
equivalent diagrams shown in Figs. 3(c), 3(d) and 3(e). So that it can
be noted in either way,
\begin{equation}
\lambda_3\left(\begin{array}{lll}p&k\\n&j\\m&i\end{array}\right)=
\lambda_3\left(\begin{array}{lll}n&i\\m&k\\p&j\end{array}\right)=
\lambda_3\left(\begin{array}{lll}m&j\\p&i\\n&k\end{array}\right)\ ,
\end{equation}
the cobosons of the lower lines, namely, $(m,i)$, $(p,j)$ or $(n,k)$, again having
the same fermion $\alpha$, by convention. These three
$\lambda_3$'s represent the same exchange process
which, according to the rules to calculate Shiva diagrams
given above, is associated to the integral
\begin{eqnarray}
\int d\v r_{\alpha_1}\,d\v r_{\alpha_2}\,d\v r_{\alpha_3}\,d\v
r_{\beta_1}\,d\v r_{\beta_2}\, d\v r_{\beta_3}\,\phi_p^\ast(\v
r_{\alpha_2},\v r_{\beta_3})\,\phi_n^\ast(\v r_{\alpha_3},\v
r_{\beta_1})\,\phi_m^\ast(\v r_{\alpha_1},\v r_{\beta_2})\nonumber
\\ \times\ \phi_i(\v r_{\alpha_1},\v r_{\beta_1})\,\phi_j(\v
r_{\alpha_2},\v r_{\beta_2})\,\phi_k(\v r_{\alpha_3},\v r_{\beta_3})\
,
\end{eqnarray}
since $m$ and $i$ have the same fermion $\alpha$ located at $\v
r_{\alpha_1}$, and so on (see Fig.4b).

(iv) \emph{In the case of $N=4$ cobosons}, one of the possible fermion
exchanges corresponds to have the cobosons $(m,i)$ (as well as $(p,j)$,
$(n,k)$ and $(q,l)$) made with the same fermion $\alpha$ and $(m,j)$
(as well as $(n,i)$, $(p,l)$ and $(q,k)$) made with the same fermion
$\beta$.  This fermion exchange between
\emph{four} cobosons can be represented by one of the \emph{four}
topologically equivalent diagrams shown in Figs. 3(f), 3(g), 3(h)
and 3(i). So that it can be noted in either way,
\begin{equation}
\lambda_4\left(\begin{array}{llll}q&l\\p&k\\n&j\\m&i\end{array}\right)=
\lambda_4\left(\begin{array}{llll}n&k\\q&i\\m&l\\p&j\end{array}\right)=
\lambda_4\left(\begin{array}{llll}p&j\\m&l\\q&i\\n&k\end{array}\right)=
\lambda_4\left(\begin{array}{llll}m&i\\n&j\\p&k\\q&l\end{array}\right)\
,
\end{equation}
the cobosons of the lower line, $(m,i)$, $(p,j)$, $(n,k)$ or
$(q,l)$, again having the same fermion $\alpha$ by convention. These four
$\lambda_4$'s represent the same exchange process which, according to
Fig.4(c) and the rules to calculate Shiva diagrams, is associated to the
integral,
\begin{eqnarray}
\int d\v r_{\alpha_1}\,d\v r_{\alpha_2}\,d\v r_{\alpha_3}\,d\v
r_{\alpha_4}\,d\v r_{\beta_1}\,d\v r_{\beta_2}\, d\v r_{\beta_3}\,d\v
r_{\beta_4}\,\phi_q^\ast(\v r_{\alpha_4},\v
r_{\beta_3})\,\phi_p^\ast(\v r_{\alpha_2},\v
r_{\beta_4})\,\phi_n^\ast(\v r_{\alpha_3},\v
r_{\beta_1})\,\phi_m^\ast(\v r_{\alpha_1},\v r_{\beta_2})\nonumber
\\ \times\ \phi_i(\v r_{\alpha_1},\v r_{\beta_1})\,\phi_j(\v
r_{\alpha_2},\v r_{\beta_2})\,\phi_k(\v r_{\alpha_3},\v
r_{\beta_3})\,\phi_l(\v r_{\alpha_4},\v r_{\beta_4})\ .
\end{eqnarray}
And so on\ldots

In the following, it will be convenient to extend the concept of
``fermion exchange'' to $N=1$ coboson, with $\lambda_1(m\ i)$
reducing to $\delta_{mi}$.

We can note that, when the number of cobosons is even, the
topologically equivalent Shiva diagrams are connected $2\times 2$, by
an up/down symmetry (as seen from diagrams of Figs. 3.(a) and 3.(b),
3.(f) and 3.(i), 3.(g) and 3.(h)): This is
\emph{a priori} possible because the lower and upper lines then are
fermion $\alpha$ lines. In the case of an odd number of cobosons
however, the upper line being a fermion $\beta$ line, such an up/down symmetry
does not exist.

We can also note that, in systems with translational invariance, due to
momentum conservation in fermion exchanges, a precise calculation of
these integrals must end by showing that the sum of coboson momenta on
each side must be equal, 
$\v Q_m+\v Q_n+\v Q_p+\cdots=\v Q_i+\v Q_j+\v Q_k+\cdots$, for these
Shiva diagrams to differ from zero.

\section{Diagrammatic expansion of
$\langle v|B_0^NB_i^\dag B_0^{\dag N-1}|v\rangle$}

The simplest scalar product of $N$-coboson states is for sure the one
in which all the cobosons are in the same state 0, \emph{i.\ e.},
$i=0$. It reduces to
$N!$ if the cobosons are taken as elementary bosons. For composite
bosons, fermion exchanges between these composite particles introduce
an additional factor, called $F_N$ in our previous works [13,14]; so
that 
\begin{equation}
\langle v|B_0^N\,B_0^{\dag
N}|v\rangle=N!\,F_N\ .
\end{equation}

We have obtained $F_N$ from the recursion relation between the $F_N$'s
derived through the recursion relation between the scalar products of
$N$, $(N-1)$, $(N-2)$,\ldots cobosons 0. Actually, this recursion
relation is quite easy to recover from the expansion of
$\langle v|B_0^NB_i^\dag B_0^{\dag N-1}|v\rangle$ by taking $i=0$ in
the end.

It is rather obvious that the scalar product of
$N$-coboson states with one coboson $i$ different from 0 must contain terms
in which the coboson
$i$ is involved in fermion exchanges with a certain amount $(P-1)$ of
the
$(N-1)$ cobosons 0 of the right, to produce $P$ of the
$N$ cobosons 0 of the left, the other $(N-P)$ cobosons 0 of this
matrix element being not involved in this exchange; so that
they simply appear as
$\langle v|B_0^{N-P}B_0^{\dag N-P}|v
\rangle$. This idea leads to the expansion shown in Fig.5(a), where the
prefactor
\begin{equation} 
A_{N}^{P}=N(N-1)\cdots(N-p+1)=\frac{(N)!}{(N-P)!}\ , 
\end{equation}
on the left, corresponds to the number of ways
to choose the $P$ cobosons 0 among $N$ which are produced in
fermion exchanges of the coboson $i$ with $(P-1)$ cobosons 0, while the factor
$A_{N-1}^{P-1}$ on the right comes from the number of ways to choose the $(P-1)$
cobosons 0 among the $(N-1)$
cobosons 0 of the right, which are involved in these fermion exchanges.
Since the scalar product of $(N-P)$ cobosons 0 is
$(N-P)!\,F_{N-P}$, this leads to [16]
\begin{eqnarray}
\langle v|B_0^NB_i^\dag B_0^{\dag N-1}|v\rangle
&=& \sum_{P=1}^N\left[(N-P)!F_{N-P}\right]A_N^P\,S_i^{(P)}\,
A^{P-1}_{N-1}\nonumber\\
&=&  N!\,\sum_{P=1}^N \frac
{(N-1)!}{(N-P)!}\,F_{N-P}\,S_i^{(P)}\ ,
\end{eqnarray}
which is shown in Fig.5(b). 

The remaining diagrams $S_i^{(P)}$
of this figure correspond to all
possible fermion exchanges between the $P$ cobosons appearing in
these diagrams, namely $P$ cobosons 0 on the left and the coboson
$i$ plus
$(P-1)$ cobosons 0 on the right, with the
constraint that the cobosons 0 are ``never alone'': Indeed, we have
already counted exchange processes in which cobosons 0 are not involved
in exchanges with $i$ when we have extracted $\langle
v|B_0^{N-P}B_0^{\dag N-P}|v
\rangle$ from the scalar product, to produce the expansion of
Fig.5(a). The number inside the circle of Fig.(5) is the total
number
$P$ of cobosons involved in these fermion exchanges. 
$S_i^{(P)}$ just corresponds to the Shiva diagram between $P$
cobosons shown in Fig.6, with an additional minus sign if the
number of exchanges is odd, as usual. So that
\begin{equation}
S_i^{(P)}=(-1)^{(P-1)}X_i^{(P)}=(-1)^{(P-1)}\lambda_P
\left(\begin{array}
{lllll}0&0\\.&.\\.&.\\.&0\\0&i\end{array}\right)\ .
\end{equation}
Note that, when all the cobosons but one are in the same state 0, the
value of the Shiva diagram does not depend on the position of the index
$i$, as easy to see from
Fig.3. Also note that, for problems with translational invariance,
due to momentum conservation in fermion exchanges, these Shiva diagrams
differ from zero for $\v Q_0=\v Q_i$ only.

All this leads to the diagrammatic expansion in fermion exchanges
of \linebreak$\langle v|B_0^NB_i^\dag B_0^{\dag N-1}|v\rangle$ shown in
Fig.7.

By setting $i=0$, we readily recover the recursion relation between the
$F_N$'s as
\begin{equation}
F_N=F_{N-1}-(N-1)F_{N-2}\,\lambda_2\left(^{0\ 0}_{0\
0}\right)+(N-1)(N-2)
F_{N-3}\,\lambda_3\left(\begin{array}{lll}0&0\\0&0\\0&0\end{array}\right)
-\cdots
\end{equation}

\section{Diagrammatic expansion of $\langle v|
B_0^NB_i^\dag B_j^\dag B_0^{\dag N-2}|v\rangle$}

We now consider the scalar product of $N$-coboson states, with two
cobosons different from 0 on the \emph{same} side. Following the 
ideas used in the preceding section, we are led to think that
this matrix element contains terms in which the cobosons $(i,j)$ are
involved in fermion exchanges with a certain amount $(P-2)$ of the
$(N-2)$ cobosons 0 of the right, to produce $P$ of the $N$ cobosons 0 of
the left, the remaining $(N-P)$ cobosons 0 of this matrix element
staying ``spectators'' in this fermion exchange, so that they
simply appear as
$\langle v|B_0^{N-P} B_0^{\dag N-P}|v\rangle=(N-P)!\,F_{N-P}$. This
gives the equation shown in Fig.8(a), which leads to 
\begin{equation}
\langle v|B_0^NB_i^\dag B_j^\dag B_0^{\dag N-2}|v\rangle=N!\,\sum
_{P=2}^N\frac{(N-2)!}{(N-P)!}\,F_{N-P}\,S_{ij}^{(P)}\ ,
\end{equation}
as made clear in Fig.8(b).
The remaining diagrams $S_{ij}^{(P)}$ correspond to all possible
fermion exchanges between $(i,j)$ and $(P-2)$ cobosons 0 which produce
$P$ cobosons 0, with the constraint that the cobosons 0 are ``never
alone''.

The first set of exchanges we can think of, is made of
connected processes, \emph{i}.\ \emph{e}., Shiva diagrams having $P$
cobosons 0 on the left and the cobosons
$(i,j)$ plus
$(P-2)$ cobosons 0 on the right. Due to topological
equivalence in Shiva diagrams, we can identify $(P-1)$ different
ones. By collecting them, we are led to introduce
\begin{equation}
X_{ij}^{(P)}=\lambda_P\left(\begin{array}
{lllllll}0&0\\.&.\\.&.\\.&.\\.&0\\0&j\\0&i\end{array}\right)+\lambda_P
\left(
\begin{array}{lllllll}0&0\\.&.\\.&.\\.&0\\.&j\\.&0\\0&i\end{array}
\right)+\lambda_P\left(\begin{array}{lllllll}
0&0\\.&.\\.&0\\.&j\\.&0\\.&0\\0&i\end{array}\right)\ +\cdots\ ,
\end{equation}
with $i$ staying at the right bottom and $j$ running to any other
of the $(P-1)$ levels --- or the reverse. This set of
exchanges is shown in Fig.9. Note that the previous fermion exchange $X_i^{(P)}$ is
related to $X_{ij}^{(P)}$ through
$X_{i0}^{(P)}=(P-1)X_i^{(P)}$.

In $X_{ij}^{(P)}$, the fermions of the cobosons $(i,j)$ are connected, either
directly or through the fermions of cobosons 0, so that they can have
1 or 0 common fermion. We can also think of exchange processes in which
the fermions of
$(i,j)$ are not connected at all, like in $X_i^{(P_1)}X_j^{(P_2)}$ with
$P_1+P_2=P$: In these disconnected diagrams, the coboson
$i$ exchanges its fermions with $(P_1-1)$ cobosons 0 to produce $P_1$
cobosons 0; and similarly for the coboson $j$. Note that, while
momentum conservation in the connected diagrams $X_{ij}^{(P)}$ only
imposes
$2\v Q_0=\v Q_i+\v Q_j$ for this diagram to differ from zero, we must have
$\v Q_0=\v Q_i=\v Q_j$ in the disconnected diagrams $X_i^{(P_1)}X_j^{(P_2)}$.

This leads to write $S_{ij}^{(P)}$ as
\begin{equation}
S_{ij}^{(P)}=(-1)^{P-1}X_{ij}^{(P)}+(-1)^{P-2}\sum_{P_1,P_2\geq
1;P_1+P_2=P} X_i^{(P_1)}\,X_j^{(P_2)}\ .
\end{equation}
The signs are again related to the parity of the number of fermion
exchanges: As disconnected diagrams made of two parts have one less
fermion exchange than the connected ones involving the same amount of
cobosons, the signs of the two terms of $S_{ij}^{(P)}$ are different.

In Fig.10, we have explicitly shown the exchange diagrams
$S_{ij}^{(P)}$ appearing in Fig.8 for $P=(2,3,4)$ cobosons.

\section{Diagrammatic expansion of $\langle
v|B_0^NB_i^\dag B_j^\dag B_k^\dag B_0^{\dag N-3}|v\rangle$}

From a pedagogical point of view, let us consider one more scalar
product with all the cobosons on the left side in the same state 0. We
first expand this scalar product as in Fig.11(a). Due to eq.\ (14)
for the part with cobosons 0 alone, this leads to Fig.11(b).

The remaining diagrams correspond to all possible fermion exchanges
between $(i,j,k)$ and $(P-3)$ cobosons 0 which produce $P$ cobosons 0,
with the cobosons 0 ``never alone''. We can think of
connected exchange diagrams, $X_{ijk}^{(P)}$, made of
$(P-1)(P-2)$ Shiva diagrams
$\lambda_P$ with $i$ at the bottom right and $(j,k)$ running to any
other possible levels.

We can also think of disconnected exchange diagrams, like 
$X_i^{(P_1)}X_{jk}^{(P_2)}$, with $P_1+P_2=P$, or even like
$X_i^{(P_1)}X_j^{(P_2)}X_k ^{(P_3)}$, with $P_1+P_2+P_3=P$. Note that,
while the connected diagrams included in $X_{ijk} ^{(P)}$ differ from zero for
$3\v Q_0=\v Q_i+\v Q_j+\v Q_k$, the disconnected diagrams made of two
parts, in addition impose one of the three $(\v Q_i,\v Q_j,\v Q_k)$ to
be equal to $\v Q_0$, while those made of three parts impose  $\v
Q_i=\v Q_j=\v Q_k=\v Q_0$.

As an example, we have drawn, in Fig.12, the set of diagrams for all
possible fermion exchanges, in the case of $P=4$ cobosons.

\section{Diagrammatic expansion of
$\langle v|B_0^{N-1}B_mB_i^\dag B_0^{\dag N-1}|v\rangle$}

We now turn to scalar products of $N$-coboson states with one coboson
different from 0 on each side. We again extract the parts having
cobosons 0 only, as shown in Fig.13(a). This leads to the expansion of
Fig.13(b), which reads [16]
\begin{equation}
\langle v|B_0^{N-1}B_mB_i^\dag B_0^{\dag N-1}|v\rangle=(N-1)!\sum_{P=1}
^N\frac{(N-1)!}{(N-P)!}\,F_{N-P}\ _mS_i^{(P)}\ .
\end{equation}
The remaining diagrams $_mS_i^{(P)}$ correspond to all possible
exchange processes between the coboson $i$ and $(P-1)$
cobosons 0 which produce the coboson $m$ and the $(P-1)$
cobosons 0, with the cobosons 0 ``never alone''. The first set of
exchange processes we can think of, is again made of Shiva diagrams
between
$P$ cobosons, now having the coboson $i$ plus $(P-1)$ cobosons 0 on
the right, and the coboson
$m$ plus $(P-1)$ cobosons 0 on the left. Due to the topological
equivalence of these Shiva diagrams, we can identify $P$ different
ones. By collecting them, we are led to introduce
\begin{equation}
_mX_i^{(P)}=\lambda_P\left(\begin{array}
{llllll}0&0\\.&.\\.&.\\.&.\\0&0\\m&i\end{array}\right)+\lambda_P\left(
\begin{array}
{llllll}0&0\\.&.\\.&.\\0&0\\m&0\\0&i\end{array}\right)+\lambda_P\left(
\begin{array}
{llllll}0&0\\.&.\\0&0\\m&0\\0&0\\0&i\end{array}\right)+\cdots\ ,
\end{equation}
with $i$ staying at the right bottom and $m$ running to all possible
levels on the left (or the reverse). (Note that $_0X_i^{(P)}=P\,X_i
^{(P)}$). This set of exchanges is shown in Fig.14(a).

We can also think of disconnected exchange processes, like
$_mX^{(P_1)}\,X_i^{(P_2)}$, with $P_1+P_2=P$ and
$_mX^{(P_1)}=\left[X_m^{(P_1)}\right]^\ast$ : In this
disconnected exchange,
$P_1$ cobosons 0 exchange their fermions to produce the coboson $m$ plus
$(P_1-1)$ cobosons 0, while the coboson
$i$ exchanges its fermions with $(P_2-1)$ cobosons 0 to produce
$P_2$ cobosons 0.

This leads to [16]
\begin{equation}
_mS_i^{(P)}=(-1)^{P-1}\,_mX_i^{(P)}+(-1)^{P-2}\sum_{P_1,P_2\geq
1;P=P_1+P_2}
\,_mX^{(P_1)}\,X_i^{(P_2)}\ .
\end{equation}
The sign difference is again due to the fact that disconnected
processes made of two parts have one fermion exchange less than the
connected ones involving the same amount of cobosons. Also note that,
while the first term of $_mS_i^{(P)}$ imposes $\v Q_m=\v Q_i$ to differ from zero, the
second term, which needs two cobosons at least to take place,
$P\geq 2$, imposes $\v Q_m=\v Q_i=\v Q_0$.

The possible fermion exchanges in the case of $N=(1,2,3)$ cobosons are
shown in Figs. 14(b), 14(c) and 14(d). They contain disconnected
diagrams made of two parts if, not only $\v Q_m=\v Q_i$, but also $\v
Q_m=\v Q_0=\v Q_i$.

\section{Diagrammatic expansion of
$\langle v|B_0^{N-1}B_mB_i^\dag B_j^\dag B_0^{\dag N-2}|v\rangle$}

In view of the above examples, the reader most probably starts to
understand how the diagrams corresponding to $N$-coboson state scalar
products have to be drawn and calculated. Let us however give two more
examples, to secure this understanding.

In order to calculate $\langle v|B_0^{N-1}B_mB_i^\dag B_j^\dag
B_0^{\dag N-2}|v\rangle$, we again start by extracting the parts
having cobosons 0 only : This gives Fig.15(a). From it, we get the
expansion shown in Fig.15(b) which reads
\begin{equation}
\langle v|B_0^{N-1}B_mB_i^\dag B_j^\dag B_0^{\dag N-2}|v\rangle=
(N-1)!\sum_{P=2}^N\frac{(N-2)!}{(N-P)!}\,F_{N-P}\ _mS_{ij}^{(P)}\ ,
\end{equation}
where $_mS_{ij}^{(P)}$ corresponds to all fermion exchanges between
cobosons $(i,j)$ and $(P-2)$ cobosons 0 which produce the coboson $m$
plus $(P-1)$ cobosons 0.

The connected processes associated to these fermion exchanges correspond to the
Shiva diagrams between $P$ cobosons, having the coboson $m$ plus $(P-1)$
cobosons 0 on the left and the cobosons $(i,j)$ plus $(P-2)$ cobosons 0
on the right. This connected contribution is in fact made of $P(P-1)$
topologically different Shiva diagrams, with, for example, $m$ at
the left bottom and $(i,j)$ running to any possible levels on the right.
By collecting them, we are led to introduce
\begin{equation}
_mX_{ij}^{(P)}=\lambda_P\left(\begin{array}
{llllll}0&0\\.&.\\.&.\\.&.\\0&j\\m&i\end{array}\right)+\lambda_P\left(
\begin{array}
{llllll}0&0\\.&.\\.&0\\0&j\\0&0\\m&i\end{array}\right)+\cdots+
\lambda_P\left(
\begin{array}
{llllll}0&0\\.&.\\.&0\\0&j\\0&i\\m&0\end{array}\right)+\cdots\
+(i\leftrightarrow j).
\end{equation}

We can also think of disconnected diagrams, with still $m$ on the left
and
$(i,j)$ on the right. These disconnected diagrams can be made of two
parts, as in $_mX_i^{(P_1)} X_j^{(P_2)}$ or
$_mX^{(P_1)}X_{ij}^{(P_2)}$, or
three parts as in $_mX^{(P_1)}X_i^{(P_2)} X_j^{(P_3)}$.

The possible fermion exchanges entering $_mS_{ij}^{(P)}$ end by
appearing as 
\begin{eqnarray}
_mS_{ij}^{(P)}&=&(-1)^{P-1}\,_mX_{ij}^{(P)}\hspace{6cm}\nonumber
\\&+&(-1)^{P-2}\sum_{P_1,P_2\geq1;P_1+P_2=P}\left[_mX_j^{(P_2)}X_i^{(P_1)}
+(i\leftrightarrow j)\right]
\nonumber\\&+&(-1)^{P-2}\sum_{P_1\geq 1;P_2\geq
2;P_1+P_2=P}\, _mX^{(P_1)}
X_{ij}^{(P_2)}\nonumber\\
&+&(-1)^{P-3}\sum_{P_1,P_2,P_3\geq 1;P_1+P_2+P_3=P}\,
_mX^{(P_1)}X_i^{(P_2)}X_j^{(P_3)}.
\end{eqnarray}
The signs are again linked to the parity of the number of exchanges
involved.

While the first term of $_mS_{ij}^{(P)}$ must have $\v Q_m+\v Q_0=
\v Q_i+\v Q_j$ to differ from zero, the second term in addition imposes $\v Q_i=\v Q_0$
(or
$\v Q_j=\v Q_0$), the third term imposes $\v Q_m=\v Q_0$ and the last
term imposes all the three $(\v Q_m,\v Q_i,\v Q_j)$ to be equal to
$\v Q_0$. Also note that, for the last two terms of $_mS_{ij}^{(P)}$ to
exist, the total number $P$ of cobosons involved in these fermion
exchanges has to be larger than 2.

The $_mS_{ij}^{(P)}$'s for $P=(2,3)$ are shown in Fig.16.

\section{Diagrammatic expansion of $\langle v|B_0^{N-2}B_mB_nB_i^\dag
B_j^\dag  B_0^{\dag N-2}|v\rangle$}

As a last example, we now consider scalar products in which two
cobosons on each side differ from 0. We have used the expression of
this scalar product in our work on the lifetime and scattering rates
of $N$ composite excitons [18].

In order to calculate this scalar product, we first again
extract the parts only having cobosons 0. This leads to the expansion of
Figs. 17(a) and 17(b) which gives
\begin{equation}
\langle v|B_0^{N-2)}B_mB_nB_i^\dag B_j^\dag B_0^{\dag N-2}|v\rangle=
(N-2)!\sum_{P=2}^N\frac{(N-2)!}{(N-P)!}\,F_{N-P}\ _{mn}S_{ij}^{(P)}\ ,
\end{equation}
where $_{mn}S_{ij}^{(P)}$ contains all possible fermion exchanges
between $P$ cobosons, connected or not, in which the cobosons $(i,j)$
exchange their fermions with $(P-2)$ cobosons 0 to produce the cobosons
$(m,n)$ plus $(P-2)$ cobosons 0, in all possible ways, with the
cobosons 0 ``never alone''.

In $_{mn}S_{ij}^{(P)}$ enter connected diagrams which correspond to the
Shiva diagrams having the cobosons $(m,n)$ plus $(P-2)$ cobosons 0 on
the left and the cobosons $(i,j)$ plus $(P-2)$ cobosons 0 on the right.
There are $P(P-1)^2$ topologically different Shiva diagrams of this
type. We collect them into
\begin{equation}
_{mn}X_{ij}^{(P)}=\lambda_P\left(\begin{array}
{llllll}0&0\\.&.\\.&.\\0&.0\\n&j\\m&i\end{array}\right)+\lambda_P\left(
\begin{array}
{llllll}0&0\\.&.\\0&.\\n&0\\0&j\\m&i\end{array}\right)+\cdots+
\lambda_P\left(
\begin{array}
{llllll}0&0\\.&.\\.&0\\0&j\\n&0\\m&i\end{array}\right)+\cdots\ ,
\end{equation}
in which $i$ stays at the right bottom while $(m,n,j)$
run to any other positions. Figure 18 shows the 12
diagrams making $_{mn}X_{ij}^{(P)}$ for $P=3$.

In addition to these connected diagrams, there are also disconnected
diagrams which correspond to all possible products of fermion exchanges
with $(m,n)$ on the left and $(i,j)$ on the right. For disconnected
diagrams made of two parts, we can think of terms like
$_{mn}X_i^{(P_1)}X_j^{(P_2)}$, or $_{mn}X^{(P_1)}X_{ij}^{(P_2)}$ with
$_{mn}X^{(P_1)}=\left[X_{mn}^{(P_1)}\right]^\ast$,
or even $_mX_i^{(P_1)}\,_nX_j^{(P_2)}$. For disconnected diagrams made
of three parts, we can think of terms like $_{mn}X^{(P_1)}X_i^{(P_2)}
X_j^{(P_3)}$ or $_mX_i^{(P_1)}\,_nX^{(P_2)}X_j^{(P_3)}$ etc\ldots We can
also have disconnected diagrams made of four parts which correspond to
terms like $_mX^{(P_1)}\,_nX^{(P_2)}X_i^{(P_3)}X_j^{(P_4)}$.

All this leads to write $_{mn}S_{ij}^{(P)}$ as 
\begin{eqnarray}
& &_{mn}S_{ij}^{(P)}=(-1)^{P-1}\
_{mn}X_{ij}^{(P)}\hspace{12cm}\nonumber\\
&+& (-1)^{P-2}\sum_{P_1,P_2\geq
1;P_1+P_2=P\geq 2}\left[_mX_i^{(P_1)}\,_nX_j^{(P_2)} +(i\leftrightarrow
j)\right]\nonumber\\
&+&(-1)^{P-2}\sum_{P_1\geq
2,P_2\geq 1;P_1+P_2=P\geq 3}\left\{\left[
_{mn}X_i^{(P_1)}X_j^{(P_2)}+(i\leftrightarrow j)\right]+\left[
_mX^{(P_2)}\,_nX_{ij}^{(P_1)}+(m\leftrightarrow n)\right]\right\}
\nonumber\\
 &+&(-1)^{P-2}\sum_{P_1,P_2\geq
2;P_1+P_2=P\geq 4}\,_{mn}X^{(P_1)}X_{ij}^{(P_2)}\nonumber\\
&+&(-1)^{P-3}\sum_{P_1,P_2,P_3\geq 1;P_1+P_2+P_3=P\geq 3}\left\{\left[
_mX_i^{(P_1)}\,_nX^{(P_2)}X_j^{(P_3)}+(i\leftrightarrow j)\right]
+(m\leftrightarrow n)\right\}\nonumber\\
&+&(-1)^{P-3}\sum_{P_1\geq 2;P_2,P_3\geq
1;P_1+P_2+P_3=P\geq 4}\left[
_{mn}X^{(P_1)}X_i^{(P_2)}X_j^{(P_3)}+\,_mX^{(P_2)}\,_nX^{(P_3)}X_{ij}
^{(P_1)}\right]\hspace{3cm}\nonumber\\
&+&(-1)^{P-4}\sum_{P_1,P_2,P_3,P_4\geq 1;P_1+P_2+P_3+P_4=P\geq 4}
\,_mX^{(P_1)}\,_nX^{(P_2)}X_i^{(P_3)}X_j^{(P_4)}\ .
\end{eqnarray}
The signs are again linked to the number of fermion exchanges. Note
that, while $P\geq 2$ is imposed by the matrix element at hand, we must
have $P\geq 3$ for the $3^{rd}$
and
$5^{th}$ terms of $_{mn}S_{ij}^{(P)}$ to exist, and $P\geq 4$ for
the
$4^{th}$, $6^{th}$ and
$7^{th}$ terms to exist.

The momentum conservations imposed by the Shiva diagrams appearing
in the various $X$'s of $_{mn}S_{ij}^{(P)}$ show that we only need $\v
Q_m+\v Q_n=\v Q_i+\v Q_j$ for the first term to differ from zero. 
The second term imposes $\v
Q_m=\v Q_i$ and $\v Q_n=\v Q_j$ with $i$ possibly changed into $j$.
For the third term, we must have, in addition to $\v
Q_m+\v Q_n=\v Q_i+\v Q_j$, one of the four $(\v
Q_m,\v Q_n,\v Q_i,\v Q_j)$ equal to $\v Q_0$. The  fourth term imposes
$\v Q_m+\v Q_n=2\v Q_0=\v Q_i+\v Q_j$. The fifth term imposes 
$\v Q_m=\v Q_i$ and $\v Q_n=\v Q_0=\v Q_j$ or any of its
permutations. In the sixth term, we must have $\v
Q_i=\v Q_j=\v Q_0$ and $\v Q_m+\v Q_n=2\v Q_0$ or
$\v Q_m=\v Q_n=\v Q_0$ and $\v Q_i+\v Q_j=2\v Q_0$, while the last term
imposes the four
$(\v Q_m,\v Q_n,\v Q_i,\v Q_j)$ to be equal to $\v Q_0$.

Fortunately, in the cases of physical interest, only a few of these
conditions on the $\v Q$'s are fulfilled, so that the number of terms
entering $_{mn}S_{ij}^{(P)}$ is considerably reduced.

\section{Rules to expand scalar products of coboson states in fermion
exchanges through Shiva diagrams}

In the previous sections, we have shown how to obtain the diagrammatic
expansion in fermion exchanges of the scalar products of a few
$N$-coboson states, with increasing complexity. In view of these
examples, we are led to state the following rules to get the expansion
of
$\langle v|B_0^{N-M}B_{m_M}\cdots B_{m_1}B_{i_1}^\dag \cdots B_{i_{M'}}
^\dag B_0^{\dag N-M'}|v\rangle$.

(i) We first extract the scalar products in which only enter
cobosons 0, \emph{i.\ e.}, scalar products like $\langle
v|B_0^{N-P}B_0^{\dag N-P}|v\rangle$ with 
$P\geq sup(M,M')$. These scalar products are equal to $(N-P)!\,F_{N-P}$, where the
$F_N$'s can be expanded through Shiva diagrams according to the
recursion relation (25) with $F_0=F_1=1$. 

(ii) We add a prefactor which corresponds to the number of ways the
$(N-P)$ cobosons 0 we have extracted can be chosen among the $(N-M)$ cobosons 0 on the
left and the $(N-M')$ cobosons 0 on the right.

(iii) The remaining parts correspond to all possible fermion exchanges
between $P$ cobosons in which explicitly appear the cobosons
$(m_1\ldots m_M)$ and $(i_1\ldots i_{M'})$ which differ from 0,
plus the cobosons 0 which have not been
extracted to get $\langle v|B_0^{N-P}B_0^{\dag N-P}|v\rangle$. In these
remaining parts, the cobosons 0 must stay ``never alone'' in the fermion
exchanges, \emph{i.\ e.}, each of them has to be connected to at least one of
the cobosons different from 0 by exchange processes.

(iv) These fermion exchanges contain processes associated to
connected diagrams \linebreak $_{m_1\cdots m_M}X_{i_1\cdots
i_{M'}}^{(P)}$ which are made of all topologically different Shiva
diagrams between $P$ cobosons, with cobosons $(m_1,\ldots, m_M)$ plus
$(P-M)$ cobosons 0 on the left and cobosons $(i_1,\ldots,i_{M'})$ plus
$(P-M')$ cobosons 0 on the right.

(v) These fermion exchanges also contain processes associated to
disconnected diagrams made of products like
$\left(_{\{m\}}X_{\{i\}}^{(P_1)}\right)\left( _{\{m'\}}X_{\{i'\}}^{(P_2)}\right)
\left(\cdots\right)$ with $P_1+P_2+\cdots=P$, where
$(\{m\},\{m'\}\ldots)$ are taken among $(m_1,\ldots, m_M)$ and
$(\{i\},\{i'\}\ldots)$ are taken among
$(i_1,\ldots,i_{M'})$, in all possible ways.

(vi) In order for these disconnected diagrams to differ from zero, in problems
with translational invariance, we must, in addition to
$\v Q_{m_1}+\cdots+\v Q_{m_M}+(P-M)\v Q_0=\v Q_{i_1}+\cdots
+\v Q_{i_{M'}}+(P-M')\v Q_0$, also have other constraints between
$\v Q_0$ and the various $\v Q_i$'s imposed by the different
$X$'s entering these disconnected diagrams. These additional constraints usually
considerably reduce the number of terms appearing in a matrix element of physical
interest.

(vii) Finally, we add a $(-1)^f$ prefactor in front of each term, where
$f$ is the number of fermion exchanges involved in the term.

(viii) The Shiva diagrams corresponding to the various
$_{\{m\}}X_{\{i\}}^{(P)}$ are then calculated along the rules given in
section 2.

The quite heavy mathematical derivation of these rules, which are based
on recursion relations between scalar products of $N$, $(N-1)$,
$(N-2)$, \ldots -coboson states will be given in an independent paper. Some of
these scalar products can already be found in refs.\ [15,16] in the case of the scalar
products of sections (3,4,6).

\section{Conclusion}

This paper deals with a quite formal, but very fundamental, aspect
of our new many-body theory for composite bosons, namely how to
calculate the scalar products of $N$-composite boson states with
all fermion exchanges included in an exact way, when the
coboson number
$N$ is large, so that there is no hope to do it in a pedestrian
way, by simply using the $2\times 2$ Pauli scatterings appearing in our theory.

Through the very ``visual'' Shiva diagrams for fermion exchanges
between $N$ cobosons we describe here, we propose a systematic procedure to derive
these scalar products, as summarized in the section 9 of this paper.
Without these diagrams, it is impossible --- or at least extremely
difficult --- to follow the physics associated to fermion exchanges in
problems dealing with $N$ interacting fermion pairs, when $N$ gets larger than 2.
Indeed, all physical quantities involved in fermion-pair many-body effects in the low
density limit end by reading in terms of these scalar products, which concentrate all the
subtilities coming from exchange processes. The contributions
to the fermion-pair many-body effects coming from interactions between
fermions appear rather simply through direct scatterings
between two coboson lines. Being very similar to Coulomb scatterings
between electrons, they are represented by diagrams very similar to the
Feynman diagrams. The conceptual difference between the many-body
physics of elementary and composite quantum particles in fact comes
from fermion exchanges. These exchanges, which appear in the scalar
products of
$N$ cobosons, are nicely visualized through
the novel Shiva diagrams we here describe.

\newpage

\begin{figure}[h]
\centerline{\scalebox{0.3}{\includegraphics{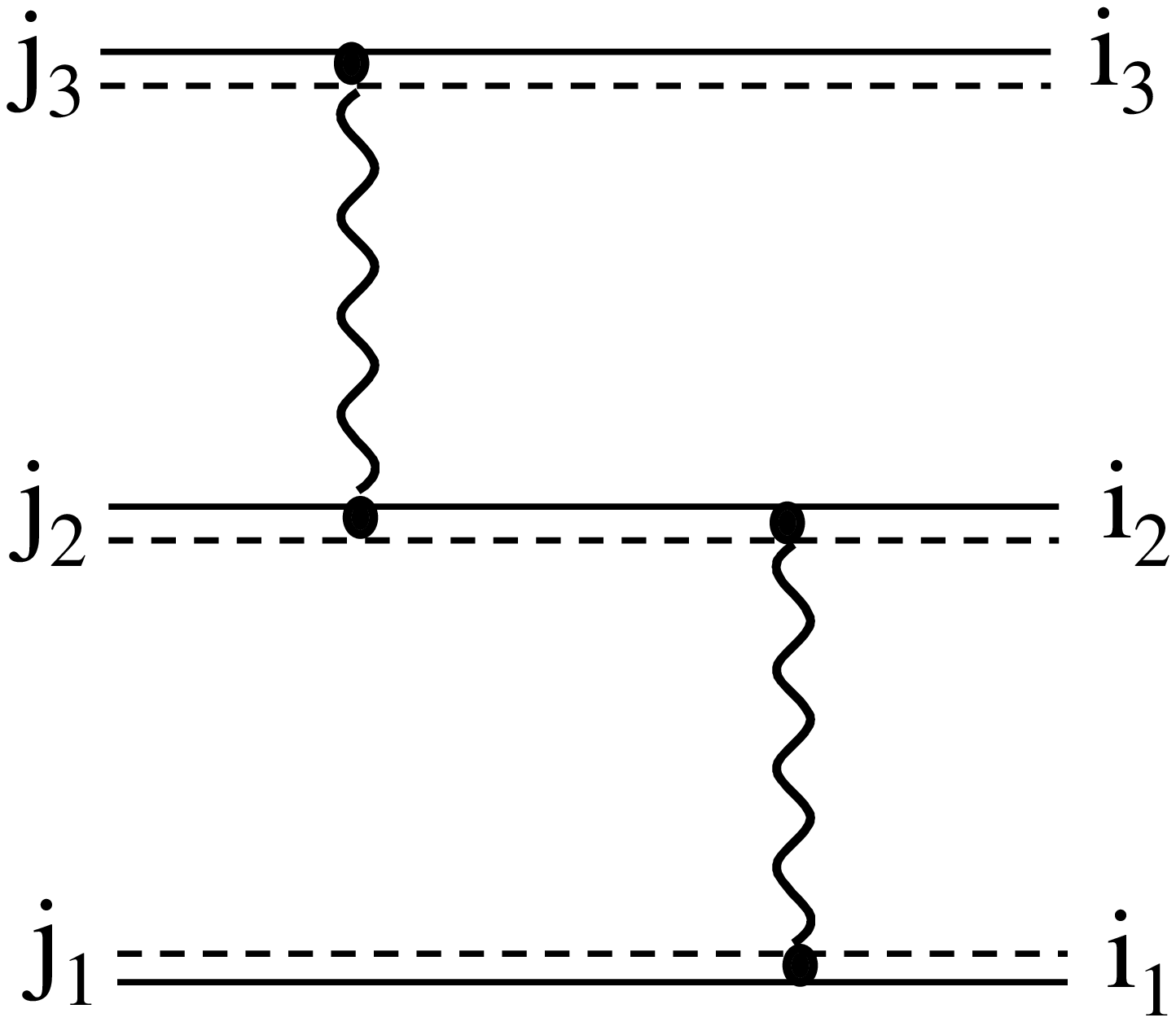}}}
\caption{A possible direct interaction between three
cobosons, starting in states $(i_1,i_2,i_3)$ and ending in states $(j_1,j_2,j_3)$. As
in all the diagrams of this paper, the fermions $\alpha$ are represented by solid lines
and the fermions $\beta$ by dashed lines.}
\end{figure}

\clearpage

\begin{figure}[h]
\centerline{\scalebox{0.7}{\includegraphics{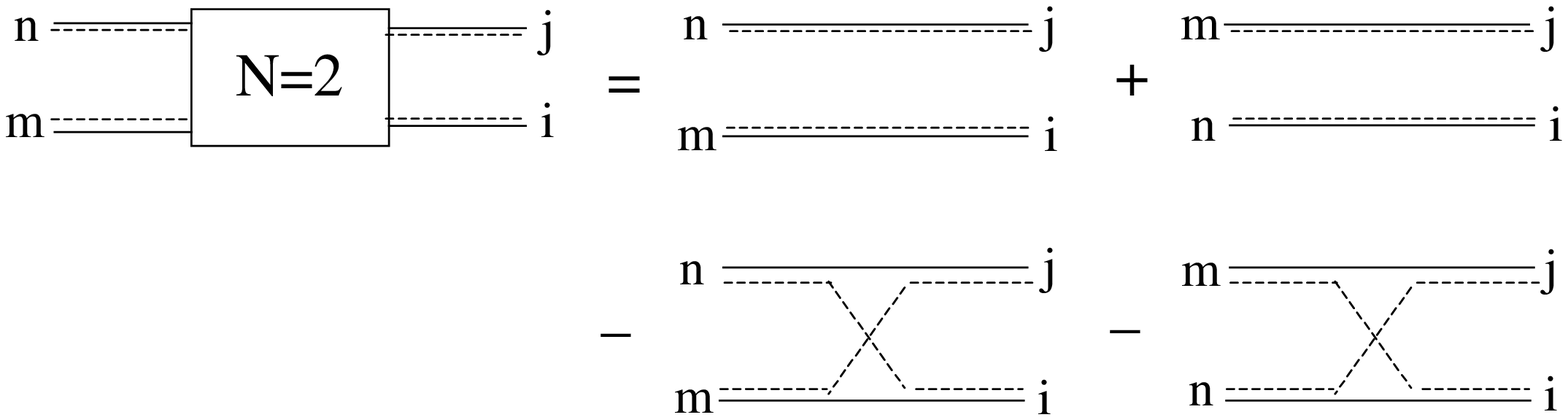}}}
\caption{Scalar product of two cobosons given in eq.\ (12). It
contains two sets of $\delta$ terms which correspond to the two upper
diagrams. These $\delta$ terms also exist for elementary bosons. This two-coboson
scalar product also contains two exchange terms which are missed when cobosons are
replaced by elementary bosons. Note that these two exchange diagrams are related by a
$(m\leftrightarrow n)$ permutation which is equivalent to a $(i\leftrightarrow j)$
permutation.}
\end{figure}
 
\clearpage

\begin{figure}[h]
\centerline{\scalebox{0.7}{\includegraphics{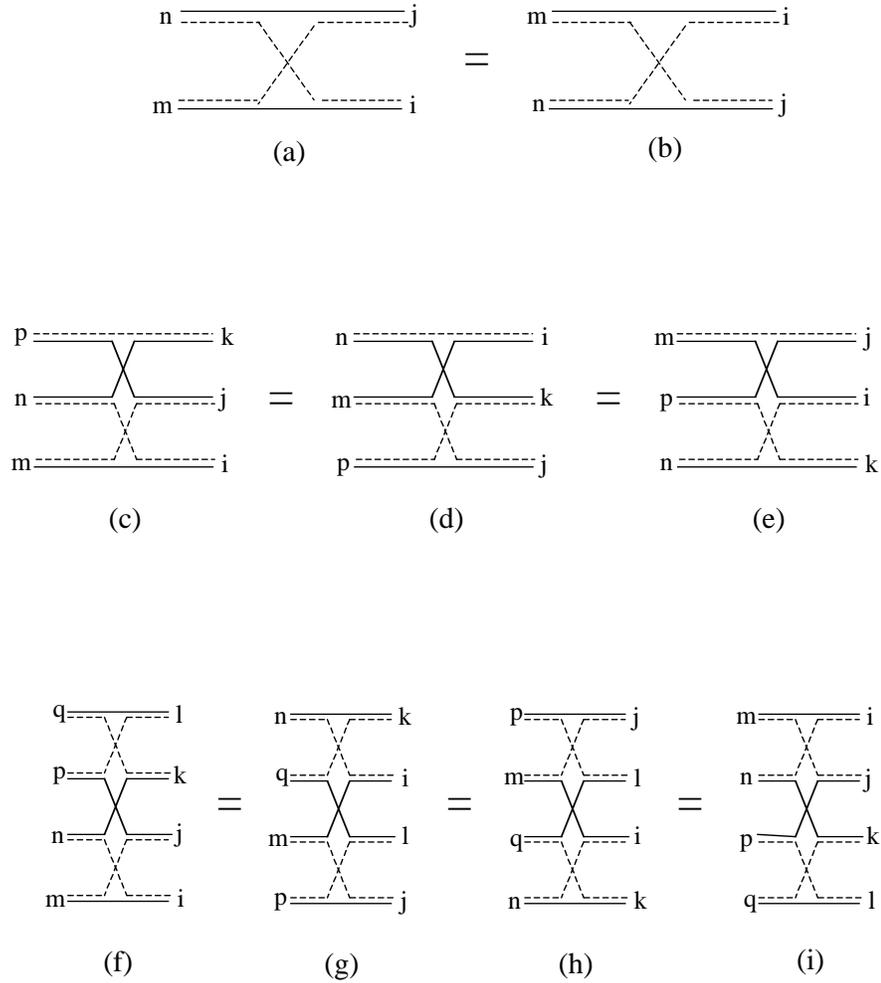}}}
\caption{(a,b): The two Shiva diagrams representing the fermion exchange between the
two cobosons $(i,j)$, in which $m$ and $i$ have the same fermion $\alpha$. (c,d,e):
The Shiva diagrams representing the fermion exchanges between the three cobosons
$(i,j,k)$, in which
$(m,i)$, $(p,j)$ and $(n,k)$ have the same fermion $\alpha$. (f,g,h,i): The four Shiva
diagrams representing the fermion exchanges between the four cobosons $(i,j,k,l)$, in
which
$(m,i)$,
$(p,j)$, $(n,k)$ and $(q,l)$ have the same fermion $\alpha$.}
\end{figure}
 
\clearpage

\begin{figure}[h]
\centerline{\scalebox{0.7}{\includegraphics{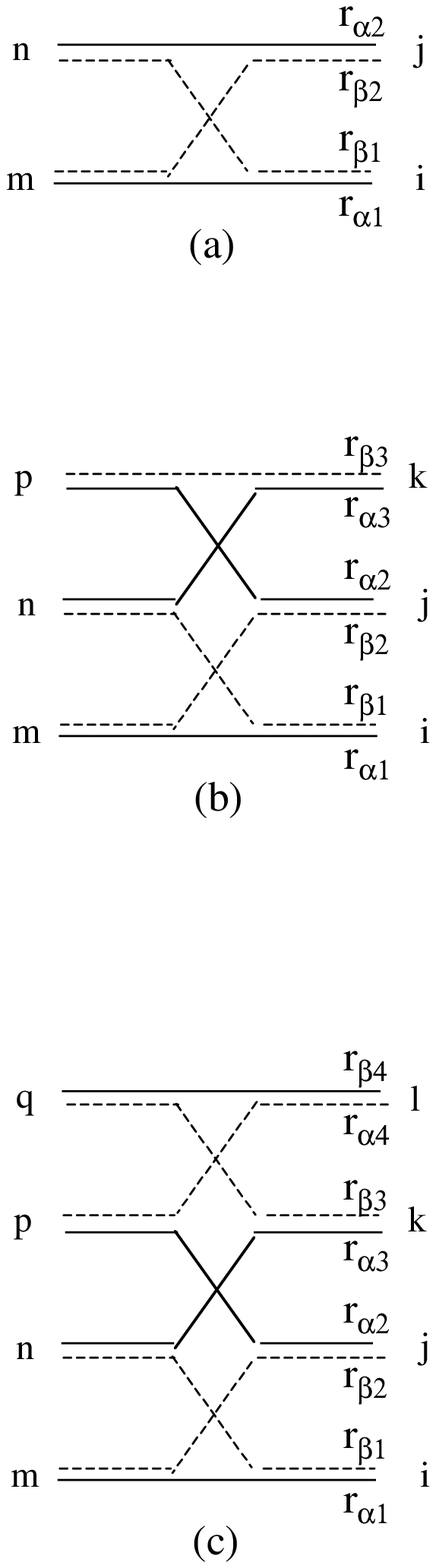}}}
\caption{Shiva diagrams between 2,3 and 4 cobosons corresponding to
the integrals given in eqs.\ (11,18,20).}
\end{figure}
 
\clearpage

\begin{figure}[h]
\centerline{\scalebox{0.7}{\includegraphics{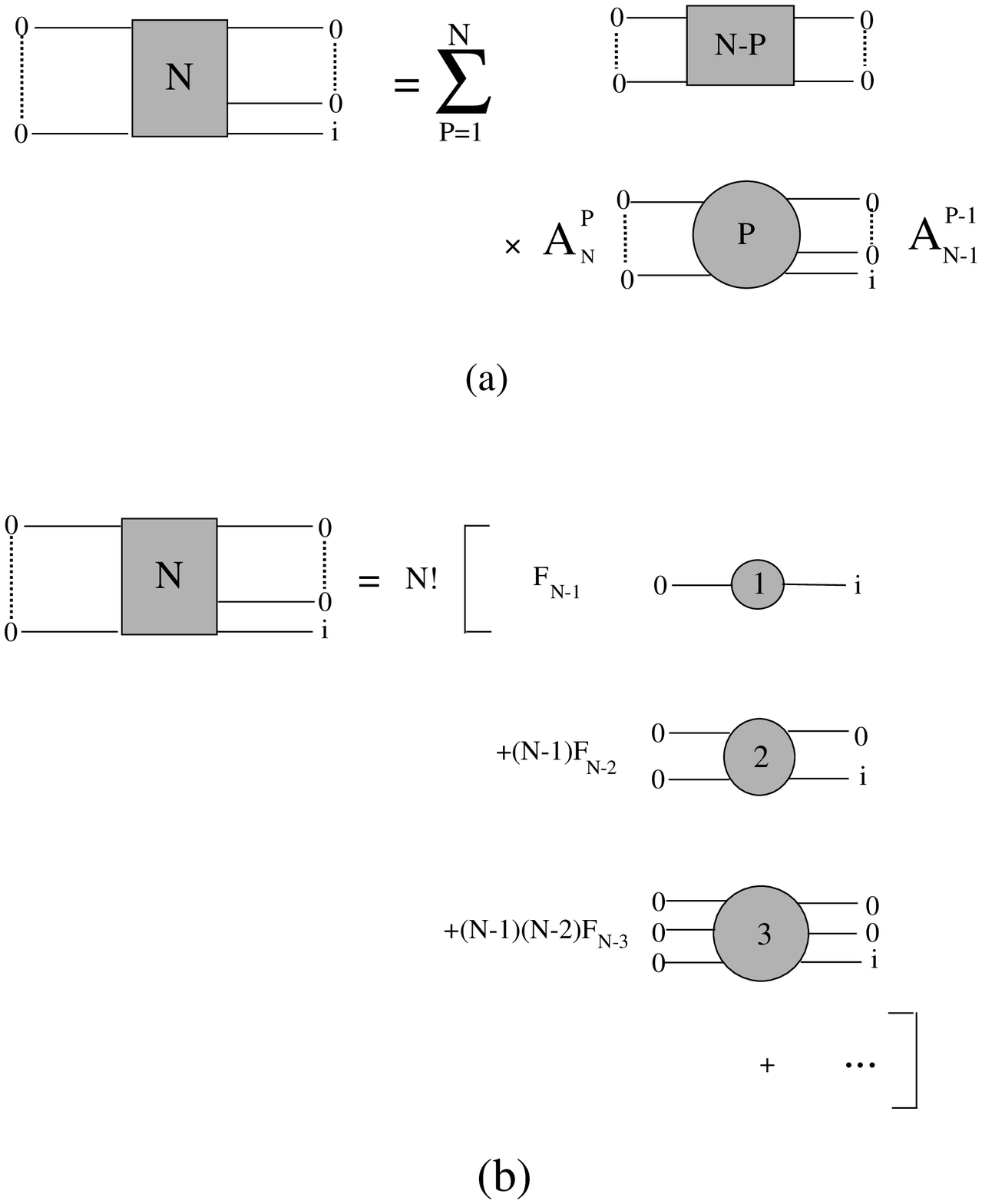}}}
\caption{Expansion of the scalar product $\langle v|B_0^NB_i^\dag B_0^{\dag
N-1}|v\rangle$ in terms of $\langle v|B_0^{N-P}B_0^{\dag
N-P}|v\rangle$, represented by the diagram with 0 cobosons only, as given in (a), and
in terms of
$F_{N-P}$ as given in (b), (see eq.\ (24)). To go from (a) to (b), we just use $\langle
v|B_{0}^{N-P}B_0^{\dag N-P}|v\rangle=(N-P)!F_{N-P}$ and the value of $A_N^P$ given in
eq.\ (22), which corresponds to the number of ways to choose the $P$ cobosons 0 of the
left, among $N$.}
\end{figure}
 
\clearpage

\begin{figure}[h]
\centerline{\scalebox{0.7}{\includegraphics{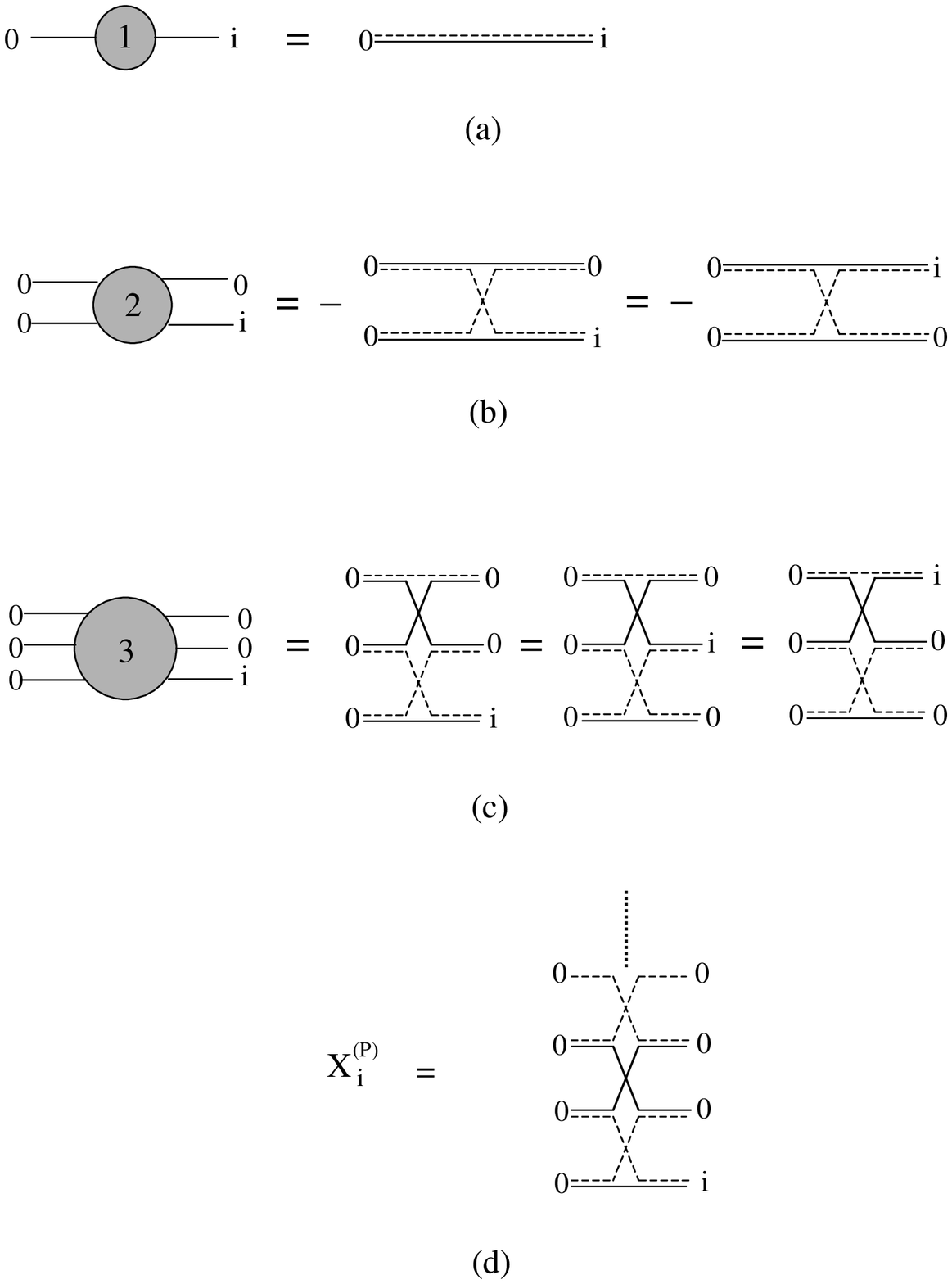}}}
\caption{(a,b,c): Fermion exchanges between coboson $i$ and $(P-1)$
cobosons 0, for $P=1,2,3$ respectively. (d): The factor
$X_i^{(P)}$ defined in eq.\ (24), which corresponds to the Shiva diagram with $P$
cobosons 0 on the left and the coboson $i$ plus $(P-1)$ cobosons 0 on the right..}
\end{figure}
 
\clearpage

\begin{figure}[h]
\centerline{\scalebox{0.7}{\includegraphics{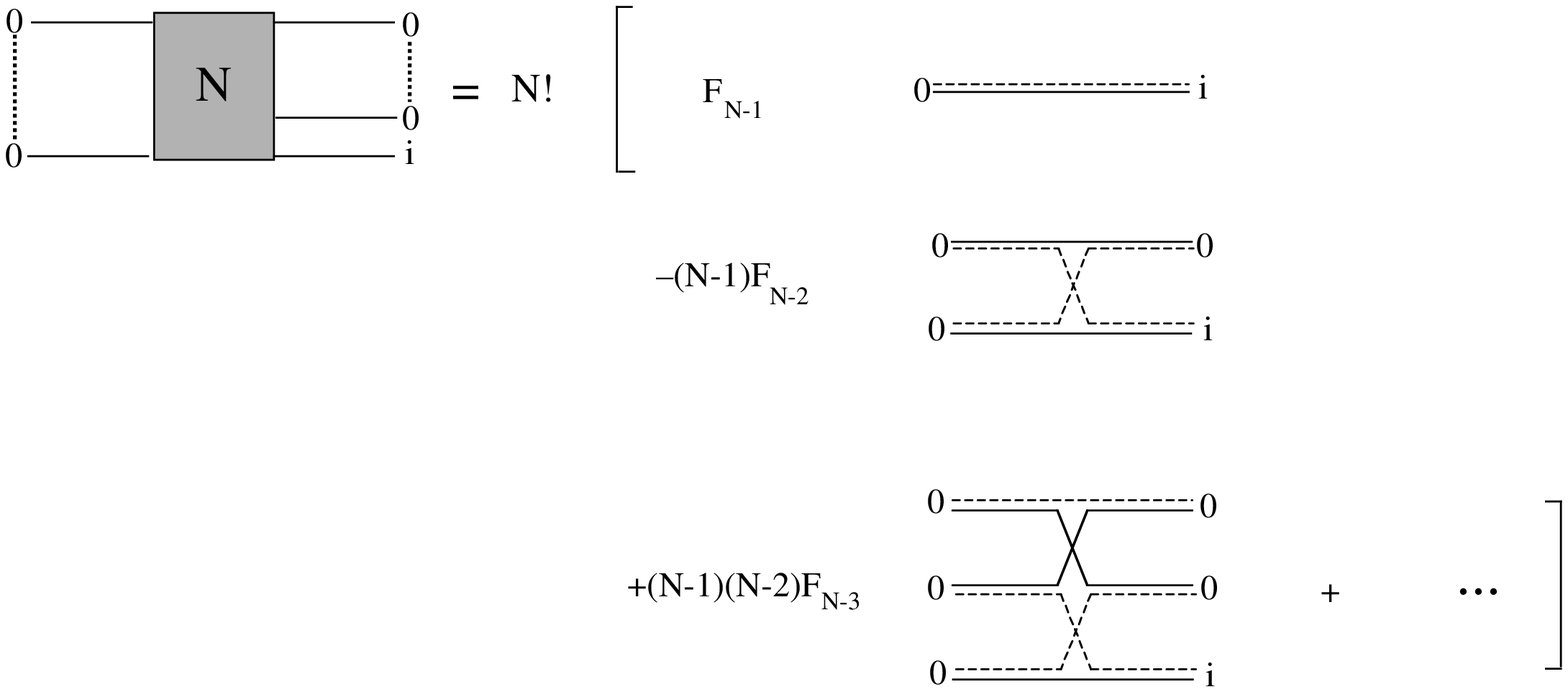}}}
\caption{Expansion of $\langle v|B_0^NB_i^\dag B_0^{\dag
N-1}|v\rangle$ in Shiva diagrams}
\end{figure}
 
\clearpage

\begin{figure}[h]
\centerline{\scalebox{0.7}{\includegraphics{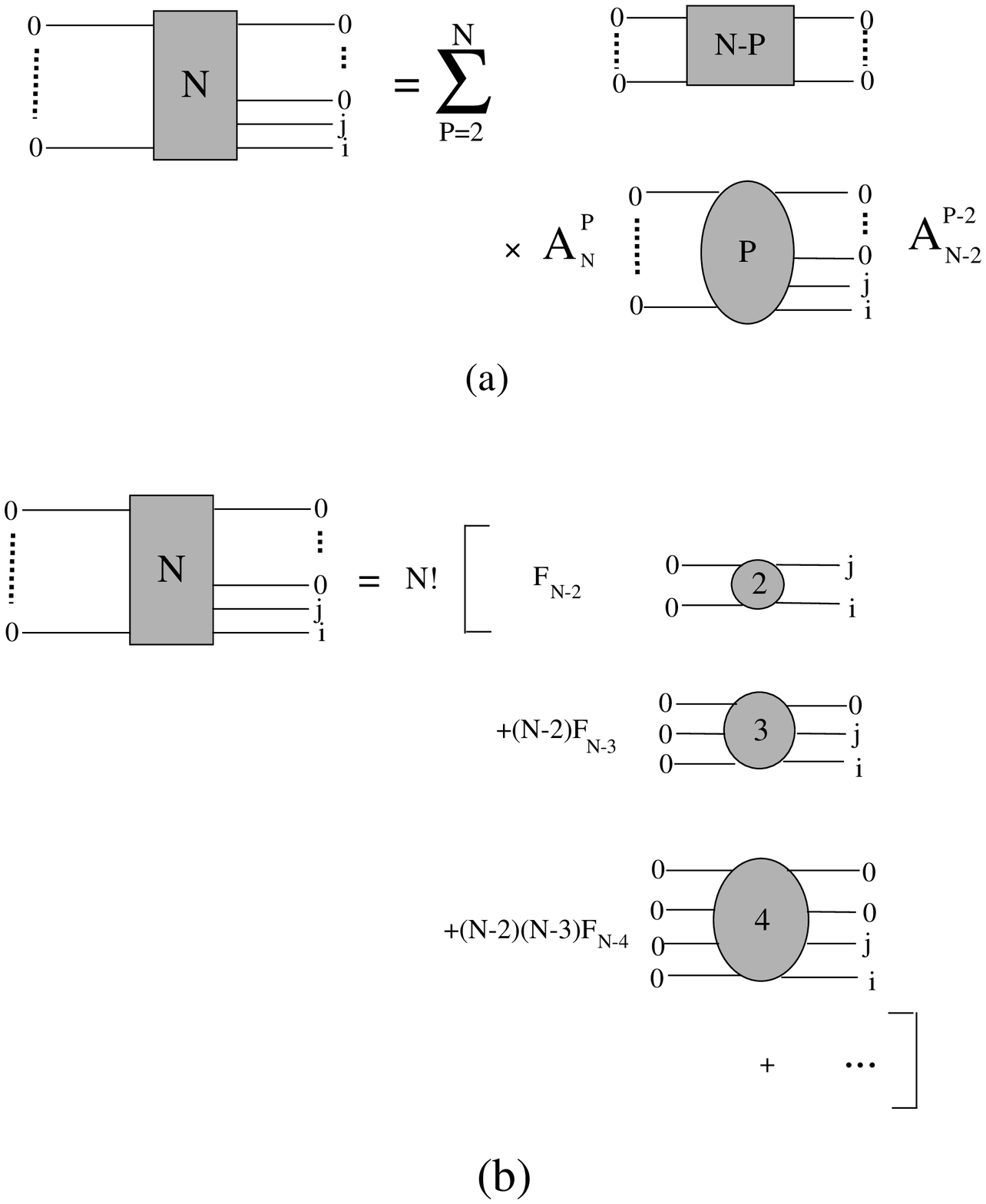}}}
\caption{Expansion of $\langle v|B_0^NB_i^\dag B_j^\dag B_0^{\dag N-2}
|v\rangle$ in terms of $\langle v|B_0^{N-P}B_0^{\dag N-P}|v\rangle$ as
given in (a), and in terms of $F_{N-P}$ as given in (b), (see eq.\
(26)).The way to go from (a) to (b) is the same as in Fig.5.}
\end{figure}
 
\clearpage

\begin{figure}[h]
\centerline{\scalebox{0.7}{\includegraphics{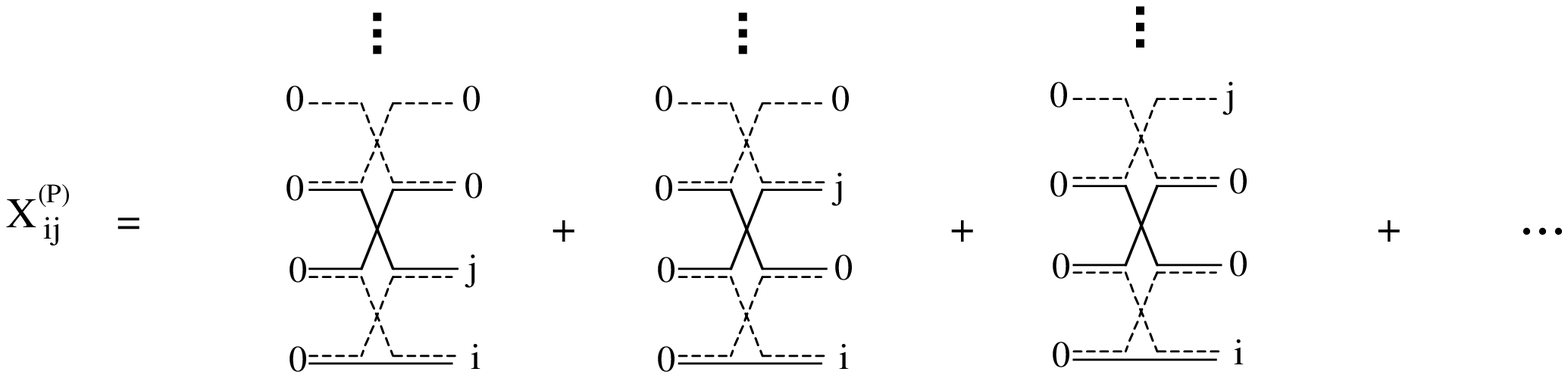}}}
\caption{Connected fermion exchanges between the cobosons $(i,j)$ and
$P-2$ cobosons 0 leading to $P$ cobosons 0.}
\end{figure}
 
\clearpage

\begin{figure}[h]
\centerline{\scalebox{0.7}{\includegraphics{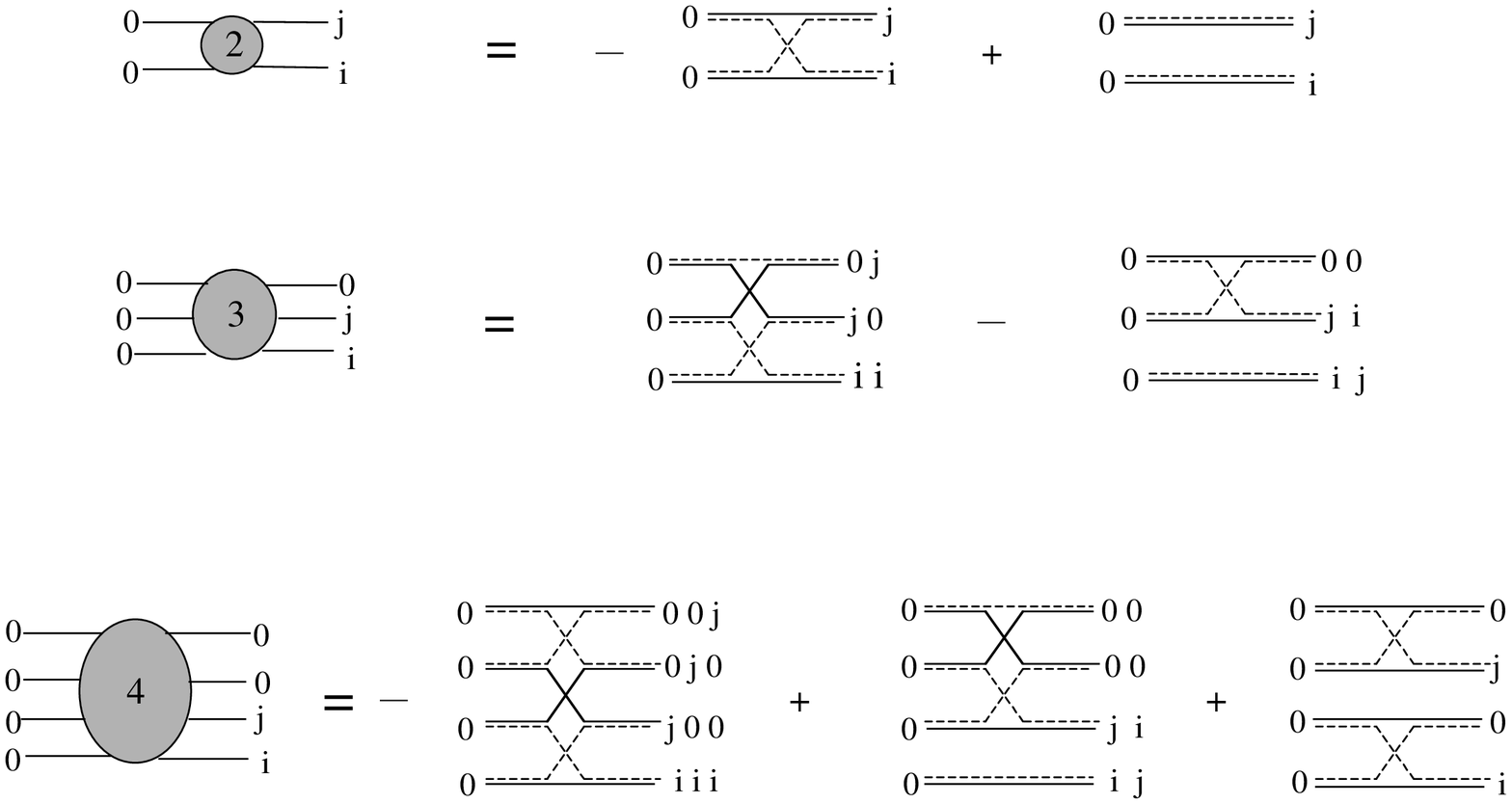}}}
\caption{Connected and disconnected fermion exchanges between the cobosons
$(i,j)$ and $P-2$ cobosons 0 leading to $P$ cobosons 0, for $P=2,3,4$. Note that in
these diagrams, the cobosons 0 are ``never alone'', \emph{i.\ e.}, they are connected to
$i$ and/or $j$. The various columns of indices, which
correspond to topologically different diagrams, have to be taken
alternatively. So that the first diagram with $P=4$ cobosons actually corresponds to 3
diagrams.}
\end{figure}
 
\clearpage

\begin{figure}[h]
\centerline{\scalebox{0.7}{\includegraphics{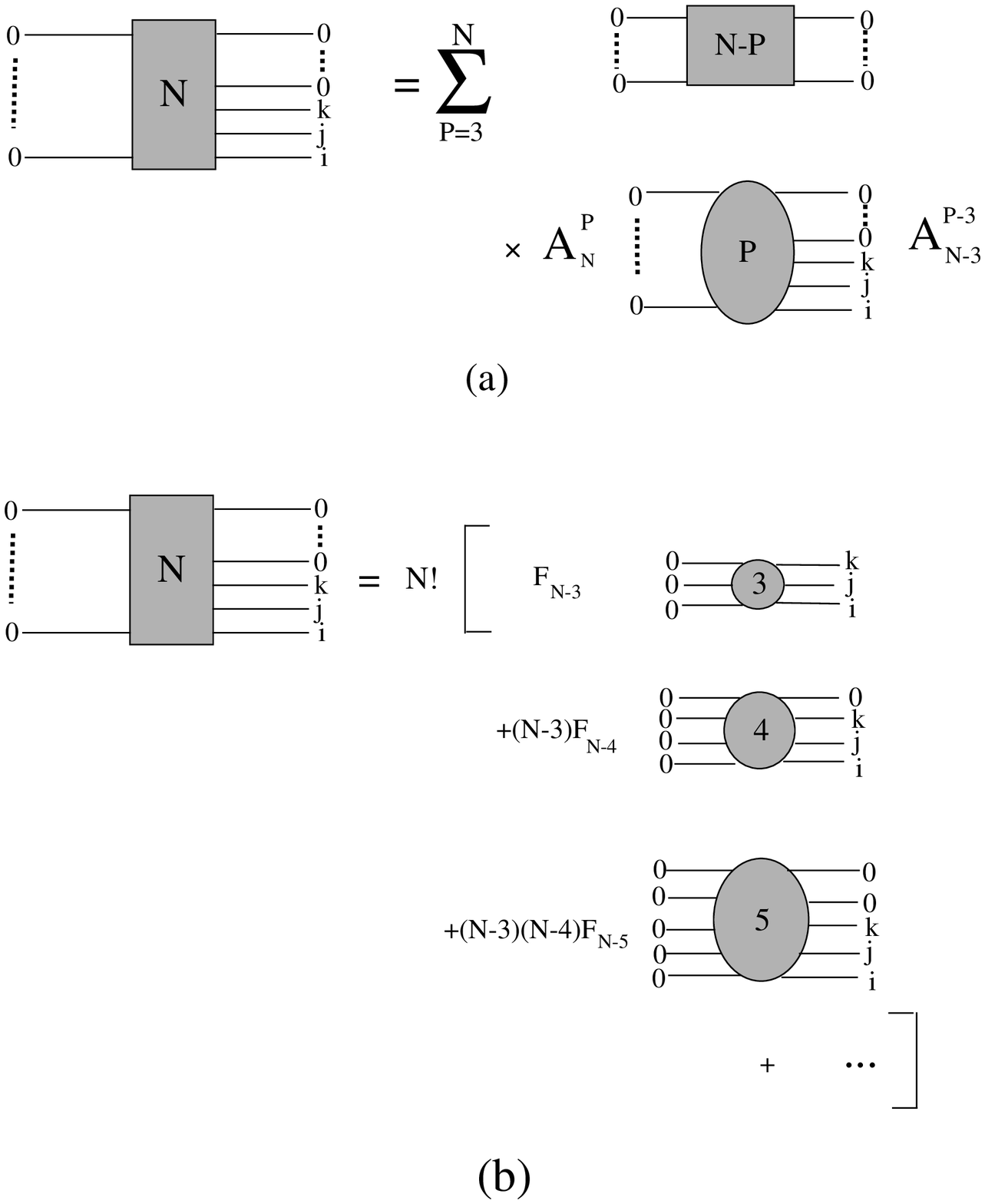}}}
\caption{Expansion of $\langle v|B_0^NB_i^\dag B_j^\dag B_k^\dag
B_0^{\dag N-3}|v\rangle$ in terms of $\langle v|B_0^{N-P}B_0^{\dag
N-P}|v \rangle$ as given in (a), and in terms of $F_{N-P}$ as given in
(b). The way to go from (a) to (b) is the same as in Fig.5.}
\end{figure}
 
\clearpage

\begin{figure}[h]
\centerline{\scalebox{0.7}{\includegraphics{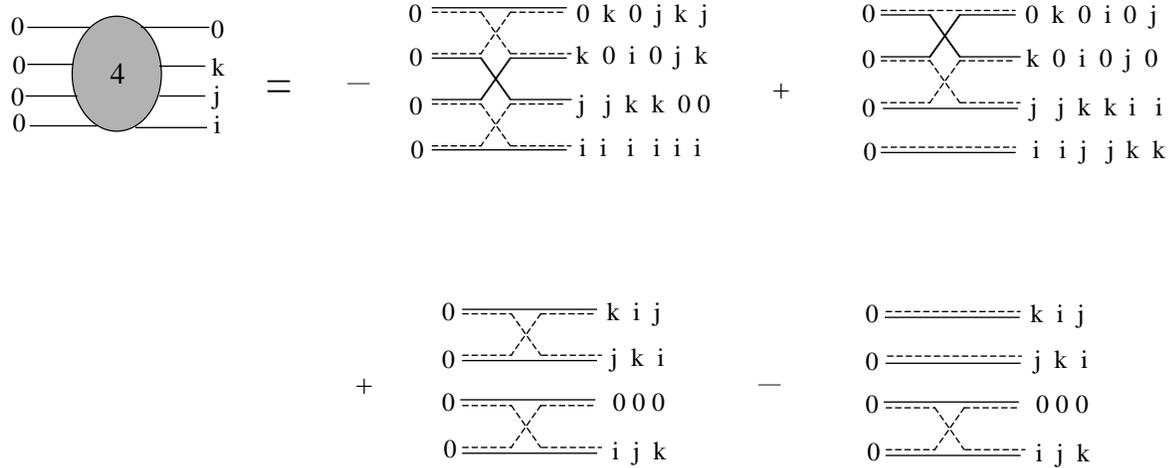}}}
\caption{Topologically different exchange processes between the cobosons
$(i,j,k)$ and one coboson 0 leading to 4 cobosons 0. The first two
diagrams appear 6 times with different index positions, while the last
two diagrams appear 3 times.}
\end{figure}
 
\clearpage

\begin{figure}[h]
\centerline{\scalebox{0.7}{\includegraphics{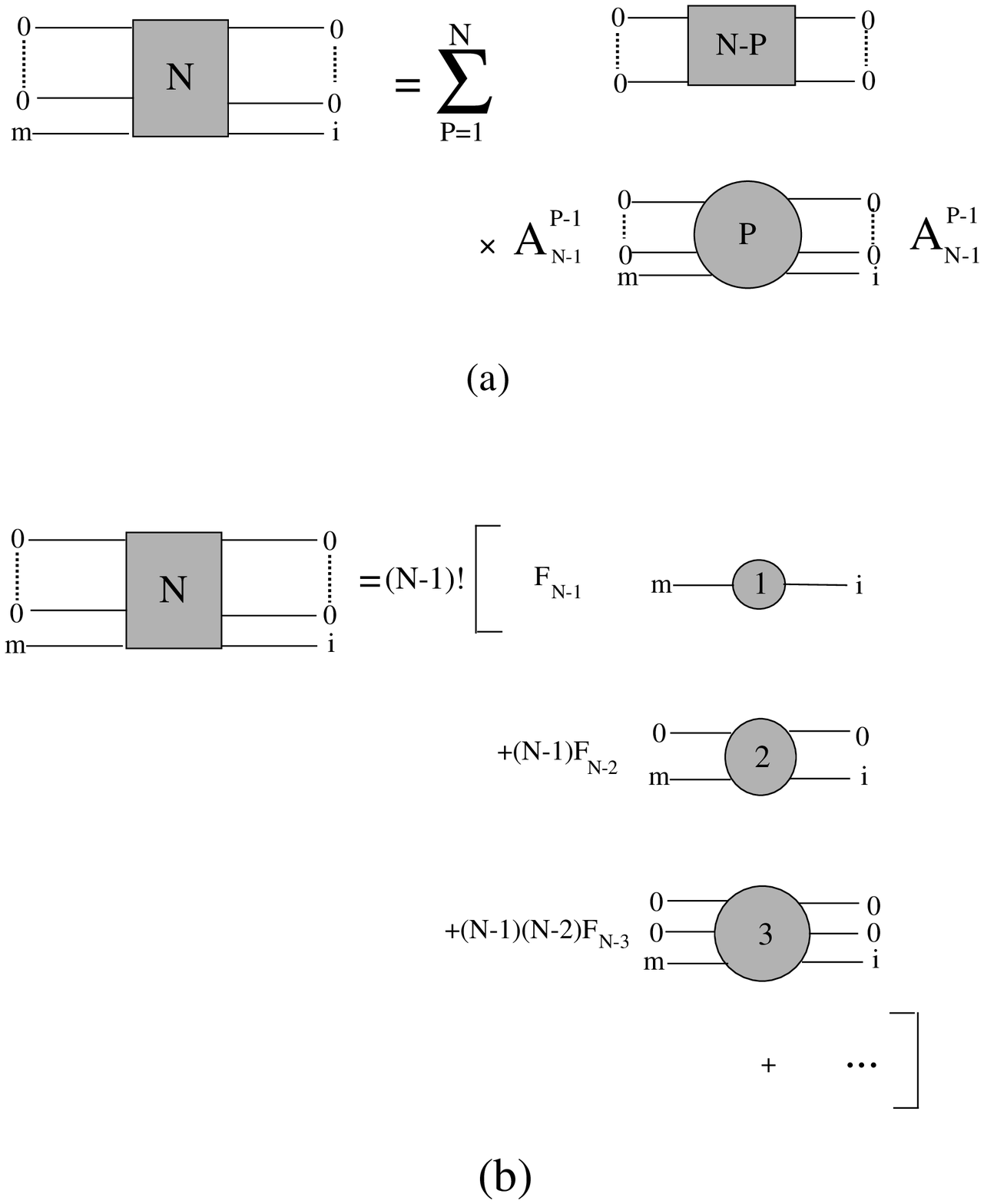}}}
\caption{Expansion of $\langle v|B_0^{N-1}B_mB_i^\dag B_0^{\dag
N-1}|v\rangle$ in terms of $\langle v|B_0^{N-P}B_0^{\dag
N-P}|v\rangle$ as given in (a), and in terms of $F_{N-P}$ as given in
(b), (see eq.\ (29)). The way to go from (a) to (b) is the same as in Fig.5.}
\end{figure}
 
\clearpage

\begin{figure}[h]
\centerline{\scalebox{0.7}{\includegraphics{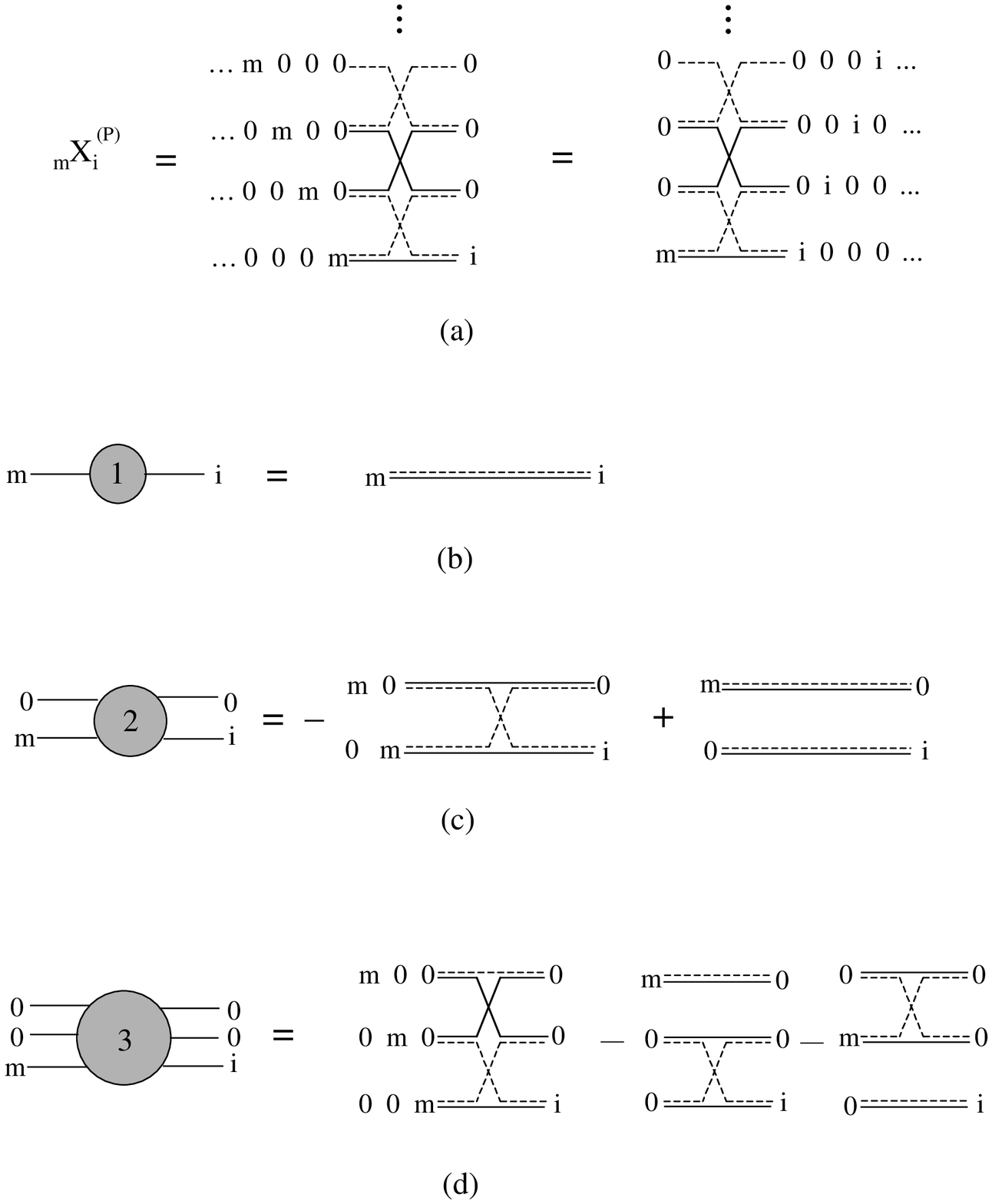}}}
\caption{(a): The different connected exchange processes between the
coboson $i$ and $(P-1)$ cobosons 0, leading to the coboson $m$ and $(P-1)$
cobosons 0. (b,c,d): Topologically different exchange processes,
connected or not, between the coboson $i$ and $(P-1)$ cobosons 0 leading to the coboson
$m$ and $(P-1)$ cobosons 0, with
$P=1,2,3$. The first Shiva diagram of fig.(d) appears 3 times.}
\end{figure}
 
\clearpage

\begin{figure}[h]
\centerline{\scalebox{0.7}{\includegraphics{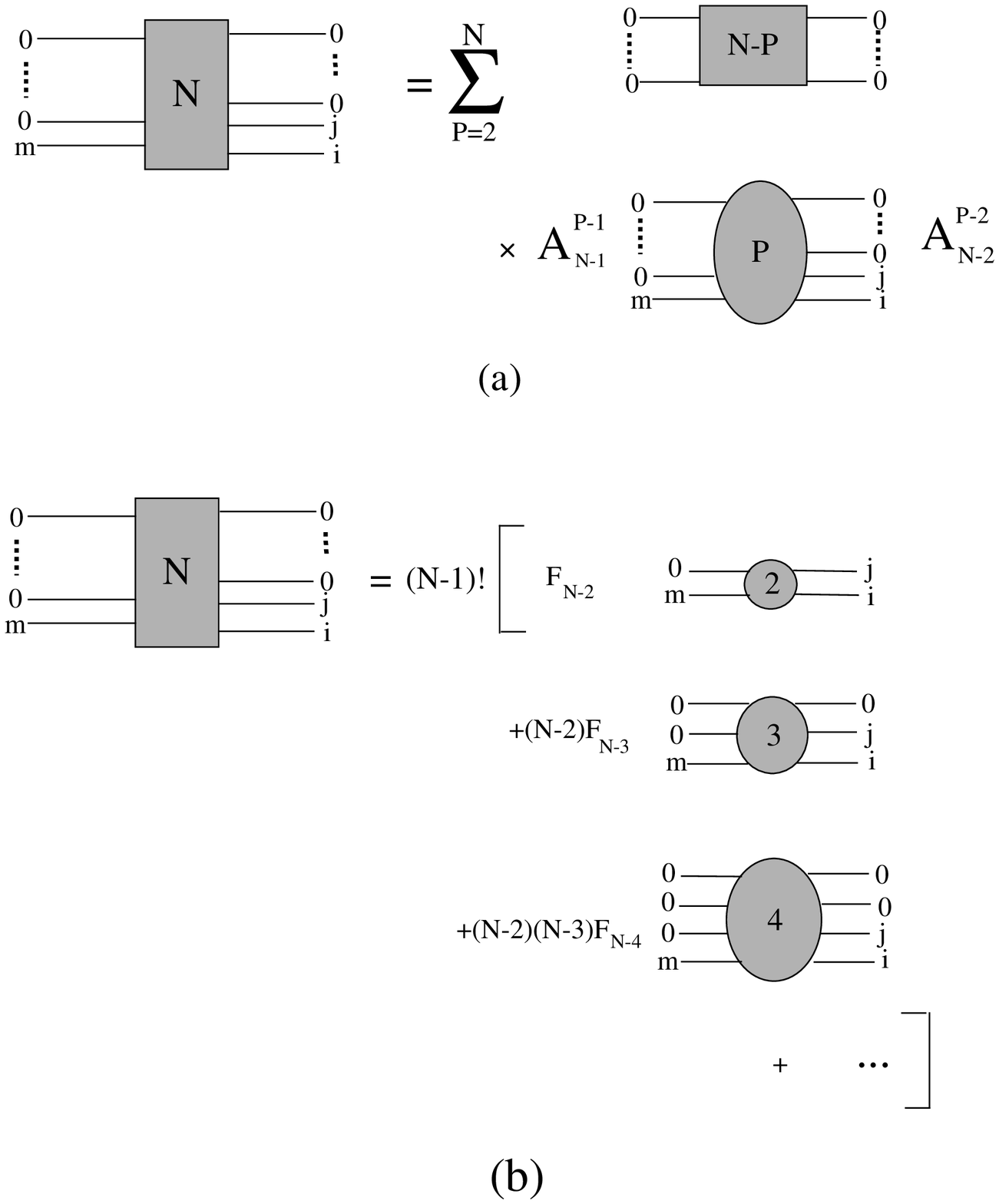}}}
\caption{Expansion of $\langle v|B_0^{N-1}B_mB_i^\dag B_j^\dag
B_0^{\dag N-2}|v\rangle$ in terms of $\langle v|B_0^{N-P}B_0^{\dag
N-P}|v\rangle$ as given in (a), and in terms of $F_{N-P}$ as given in
(b), (see eq.\ (32)). The way to go from (a) to (b) is the same as in Fig.5.}
\end{figure}
 
\clearpage

\begin{figure}[h]
\centerline{\scalebox{0.7}{\includegraphics{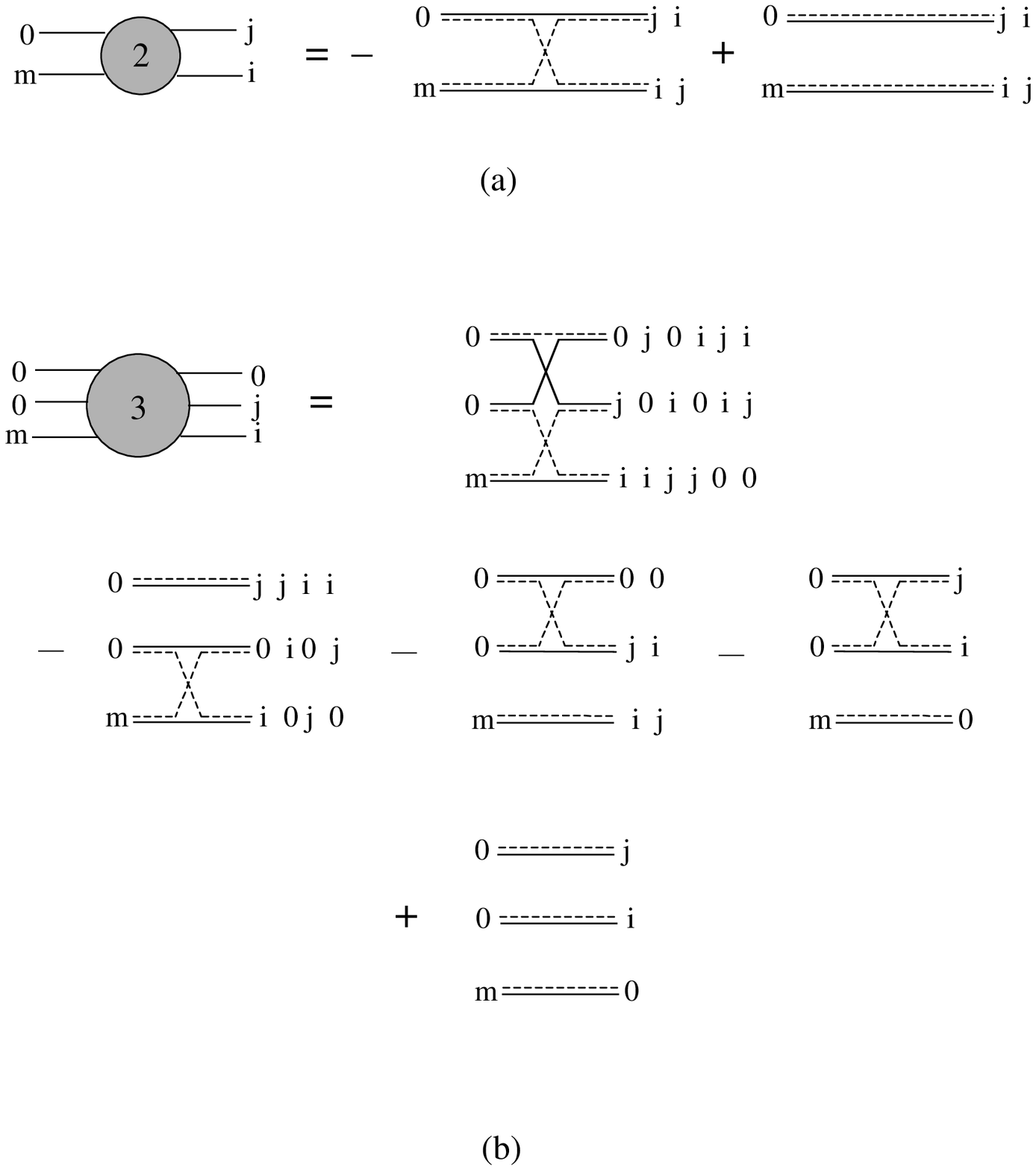}}}
\caption{Topologically different exchange processes, connected or not,
between the cobosons $(i,j)$ and $(P-2)$ cobosons 0, giving the coboson $m$ and
$(P-1)$ cobsons 0, for $P=2,3$. The first diagram of (b) appears 6 times, which
corresponds to all the possible positions of $(i,j,0)$ in this Shiva diagram. And so
on.}
\end{figure}
 
\clearpage

\begin{figure}[h]
\centerline{\scalebox{0.7}{\includegraphics{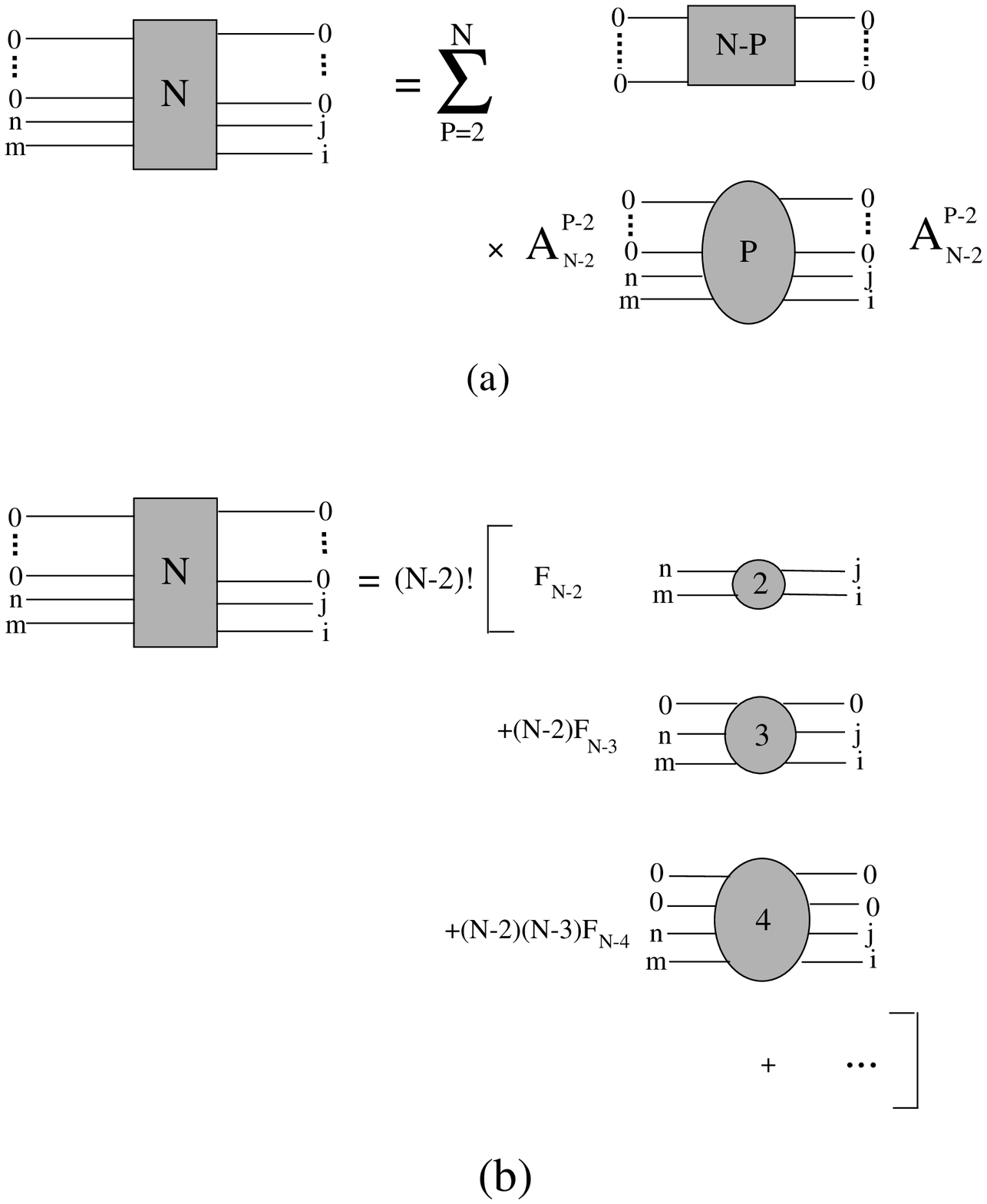}}}
\caption{Expansion of $\langle v|B_0^{N-2}B_mB_nB_i^\dag B_j^\dag
B_0^{\dag N-2}|v\rangle$ in terms of $\langle v|B_0^{N-P}B_0^{\dag
N-P}|v\rangle$ as given in (a), and in terms of $F_{N-P}$ as given in
(b), (see eq.\ (35)). The way to go from (a) to (b) is the same as in Fig.5.}
\end{figure}
 
\clearpage

\begin{figure}[h]
\centerline{\scalebox{0.3}{\includegraphics{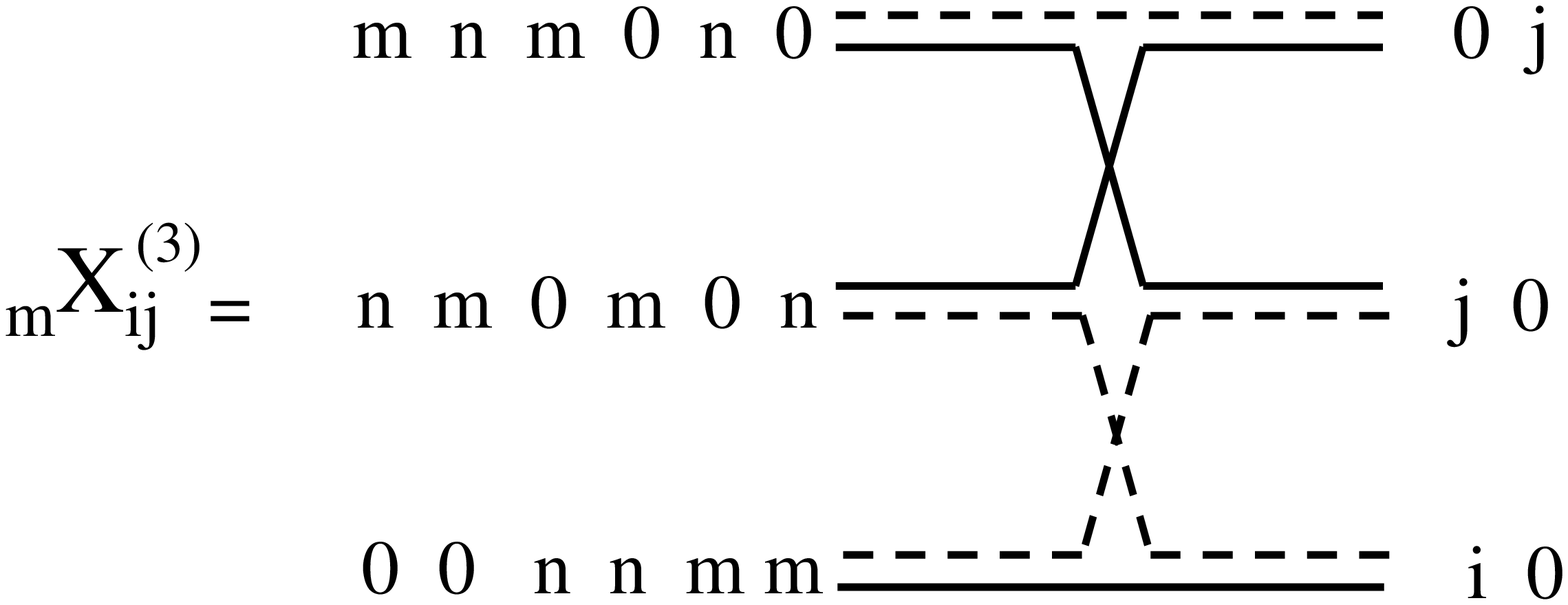}}}
\caption{The connected exchange process between the cobosons $(i,j,0)$ giving the
cobosons $(m,n,0)$. This Shiva diagram appears $(6\times 2)$ times,
due to the topologically different positions of the various indices.}
\end{figure}
 
\clearpage


\begin{thebibliography}{99}
\bibitem{11}
As explained in ref.\ [10], it is possible to deal with
non-orthogonal cobosons. For the sake of simplicity, we stay here with
orthogonal one-coboson states.
\bibitem{12}
In ref.\ [10], it is also explained how to define the creation
potential when the cobosons of interest are not the Hamiltonian
eigenstates.
\bibitem{17}
from the multiarm hindu god. In one of our very first works, we 
called them ``skeleton diagrams'', unproperly.
\end{thebibliography}
\end{document}